\documentclass[sigplan]{acmart}
\pdfoutput=1
\renewcommand\footnotetextcopyrightpermission[1]{} 

\acmISBN{} 
\acmDOI{} 
\startPage{1}

\setcopyright{none}

\usepackage{booktabs}
\usepackage{subcaption} 

\begin{document}

\title[]{Toward Code Generation: A Survey and Lessons from Semantic Parsing}

\author{Celine Lee}
\affiliation{ 
  \institution{Intel Labs, University of Pennsylvania} 
}
\email{celine.lee@intel.com}

\author{Justin Gottschlich}
\affiliation{
  \institution{Intel Labs, University of Pennsylvania} 
}
\email{justin.gottschlich@intel.com}  

\author{Dan Roth}
\affiliation{
  \institution{University of Pennsylvania}   
}
\email{danroth@seas.upenn.edu}    


\begin{abstract} 
With the growth of natural language processing techniques and demand for improved software engineering efficiency, there is an emerging interest in translating intention from human languages to programming languages. In this survey paper, we attempt to provide an overview of the growing body of research in this space. We begin by reviewing natural language semantic parsing techniques and draw parallels with program synthesis efforts. We then consider semantic parsing works from an evolutionary perspective, with specific analyses on neuro-symbolic methods, architecture, and supervision. We then analyze advancements in frameworks for semantic parsing for code generation. In closing, we present what we believe are some of the emerging open challenges in this domain.

\end{abstract}



\maketitle
\pagestyle{plain}

\section{Introduction}

\begin{figure}
\includegraphics[width=0.85\linewidth]{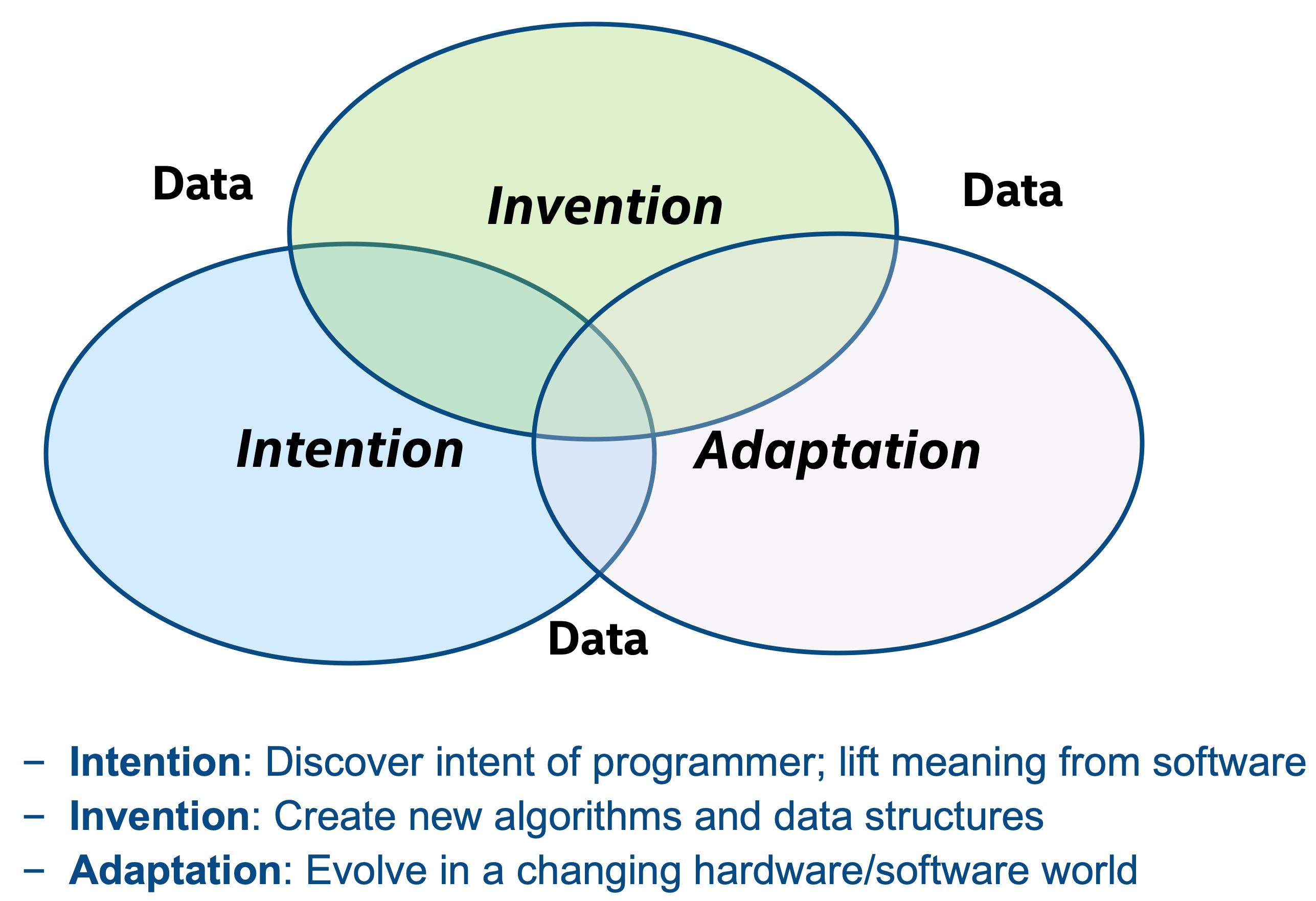}
\caption{The Three Pillars of Machine Programming (credit: Gottschlich et al.~\cite{gottschlich2018pillars}).}\label{fig:three_pillars}
\end{figure}

\emph{Machine programming} (MP) is the field concerned with the automation of all aspects of software development~\cite{gottschlich2018pillars}. According to \citet{gottschlich2018pillars}, the field can be reasoned about across three pillars: \emph{intention}, \emph{invention}, and \emph{adaptation} (Figure~\ref{fig:three_pillars}). \emph{Intention} is concerned with capturing user intent, whether by natural language, visual diagram, software code, or other techniques. \emph{Invention} explores ways to construct higher-order programs through the composition of existing -- or novel creation of -- algorithms and data structures. \emph{Adaptation} focuses on transforming higher-order program representations or legacy software programs to achieve certain characteristics (e.g., performance, security, correctness, etc.) for the desired software and hardware ecosystem.

In this paper, we discuss \emph{(i)} semantic parsing and \emph{(ii)} code generation, which chiefly fall within the boundaries of the intention and invention pillars, respectively. In the context of natural language (NL), semantic parsing is generally concerned with converting NL \emph{utterances} (i.e., the smallest unit of speech) to structured logical forms. Once done, these logical forms can then be utilized to perform various tasks such as question answering \cite{kate:2005:ncai,krishnamurthy:2017:empiricalmethodsnl,waltz:1978:acm,yih:2014:acl,liang:2011:acl}, machine translation \cite{andreas:2013:acl,wong:2006:naacl}, or code generation \cite{yin-neubig-2017-syntactic,iyer-etal-2018-mapping,ling2016latent,shin2019program}, amongst other things. 
For semantic parsing for code generation, the end logical form will usually be some form of software code. This may be the direct output program \cite{ling2016latent,zhong:2018:arxiv,kate:2005:ncai,zelle:1996:aaai} or an intermediate representation of code \cite{yin-neubig-2017-syntactic,yin:2018:emnlp,dong2016language,rabinovich:2017:acllong,krishnamurthy:2017:empiricalmethodsnl}, which can subsequently be fed into a program synthesis component to generate syntactically sound and functionally correct code.

%

In this survey paper, we first provide an overview of the processes of NL semantic parsing and of code generation in Sections~\ref{sec:semanticparsing} and \ref{sec:program-synthesis}, respectively. In Section~\ref{sec:evolution}, we summarize the evolution of techniques in NL and code semantic parsing from the 1970s onward. Section~\ref{sec:supervision} explores the question of supervision for semantic parsing tasks, while Section~\ref{sec:modern} discusses modern advances in neural semantic parsing for code generation. To conclude, we consider some possible future directions of semantic parsing for code generation.

\section{Natural Language Semantic Parsing}
\label{sec:semanticparsing}

\begin{figure}
\includegraphics[width=0.65\columnwidth]{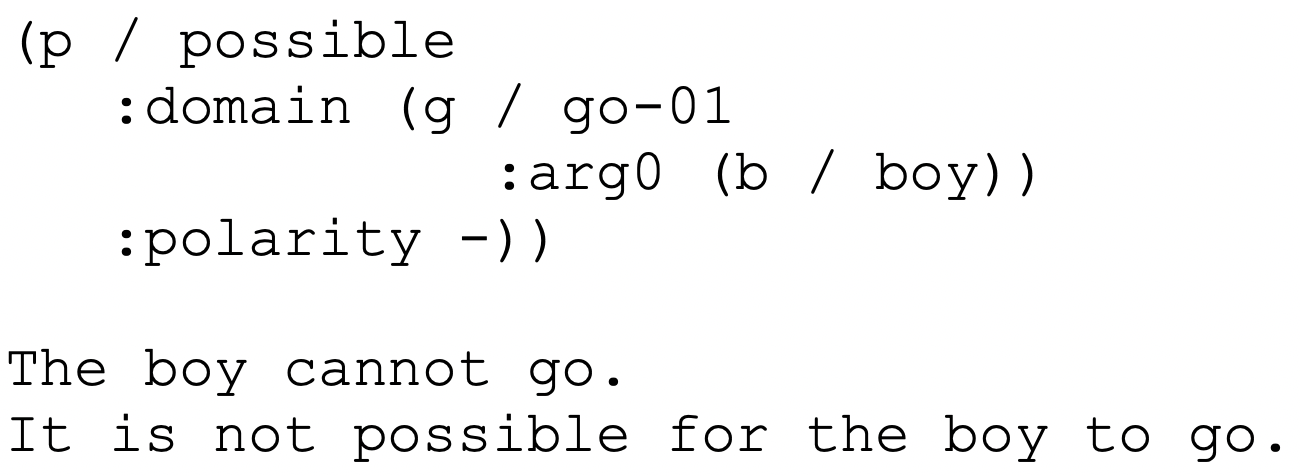}
\caption{Abstract meaning representation (AMR) of two semantically equivalent, but syntactically different sentences (credit: ~\citet{banarescu-etal-2013-abstract}).}
\label{fig:boy}
\end{figure}

Natural language analysis can be segmented into at least two fundamental categories: \emph{syntax} and \emph{semantics}. In general, \emph{syntax} is the way a natural language phrase is structured; the order and arrangement of words and punctuation. The \emph{semantics} is the meaning that can be derived from such syntax. For example, the two sentences \emph{“the boy cannot go”} and \emph{“it is not possible for the boy to go”}, are semantically equivalent even though they are syntactically different. By lifting semantics from syntax, NL semantic parsers can map semantically equivalent sentences to the same logical form, even when they have different syntaxes. An example of this is shown in Figure~\ref{fig:boy}, where two syntactically different sentences are shown to be semantically equivalent using an abstract meaning representation (AMR) as its logical form (more details in Section~\ref{sec:meaningrep}). For the remainder of this section, we discuss how NL semantic parsers extract meaning from natural language \emph{utterances} into various machine-understandable representations.

\subsection{Purpose of Natural Language Semantic Parsing}

The general purpose of structured logical forms for natural language is to have a representation that enables natural language understanding (NLU) and automated reasoning, amongst other things. 
Historically, NL semantic parsers have enabled computers to query databases \citep{yu:2018:emnlp}, play card games \citep{goldwasser:2014:mljournal}, and even act as conversational agents such as Apple's Siri and Amazon's Alexa. Semantic parsing has taken on many forms. Early hand-crafted rules made way for supervised learning techniques. Then, due to the labor-intensive and error-prone process of hand-labeling data for paired natural language and logical form datasets, researchers turned their attention toward weakly supervised \cite{artzi:2013:tacl,dasigi:2019:acl} and other learning techniques that tend to require less labeled data. 


\subsection{Components of a Semantic Parsing System}

Given an NL input, a NL semantic parsing system generates its semantic representation as a structured logical form. These logical forms are also known as meaning representations or programs. To generate these outputs, semantic parsing pipelines usually trained by some learning algorithm to produce a distribution over the output search space and then search for a best scoring logical form in that distribution. This logical form can then be executed against some environment to carry out an intended task (e.g. to query a database). Because these logical forms are generally designed to be machine-understandable representations, they tend to have a structured nature, following an underlying formalism by which some grammar can be used to derive valid logical forms. Insight into the grammar of a set of logical forms can be leveraged by different semantic parsing methods to improve performance. For example, semantic parsing for code generation should have a meaning representation that can deterministically be transformed into valid code. Therefore, the output space can be constrained by the underlying grammar formalisms that define syntactically sound code.


\subsection{Grammars}

\begin{figure}
    \includegraphics[width=0.7\columnwidth]{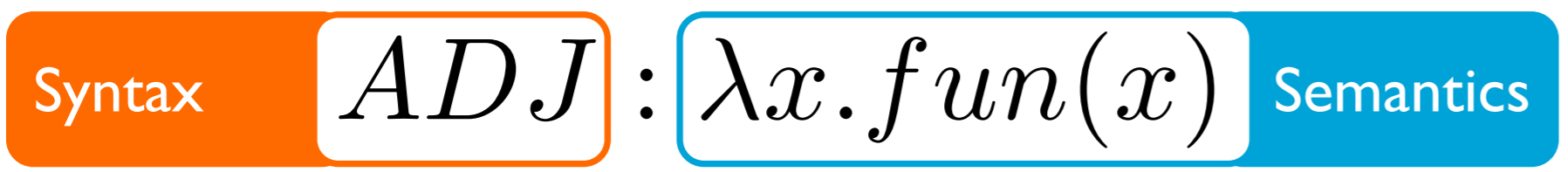}
    \caption{An example of a CCG category. Syntax (the ADJ represents an adjective part of speech) is paired with semantics (the lambda term expression) (credit: \citet{artzi-etal-2014-semantic}).}
    \label{fig:ccg}
\end{figure}

Grammar in the context of natural language processing is a set of rules that govern the formation of some structure (parsing of NL utterances) to ensure well-formedness. A familiar example is English grammar: number, tense, gender and other features must be consistent in an English sentence. The sentence \emph{“he eat apples”} is grammatically incorrect because a single noun acts with a verb that must end in ‘s.’ Likewise, in semantic parsing, a logical form can be grammatically constrained to ensure well-formedness. A set of grammatical rules constrains the search space of possible outputs. Combinatory categorical grammar (CCG) \citep{Steedman1987CombinatoryGA,Steedman2007COMBINATORYCG} is one such example of a popular grammar formalism. It has historically made frequent appearances in semantic parsing models \cite{zettlemoyer:2005:auai,kwiatkowski-etal-2013-scaling,zettlemoyer-collins-2007-online} due to its ability to jointly capture syntactic and semantic information of the constituents which it label. Consider the example in Figure~\ref{fig:ccg}. By pairing syntactic and semantic information, CCGs can describe textual input for a machine to understand the semantics with the syntax. 


\subsection{Meaning Representations}
\label{sec:meaningrep}

\begin{figure}
    \includegraphics[width=0.8\columnwidth]{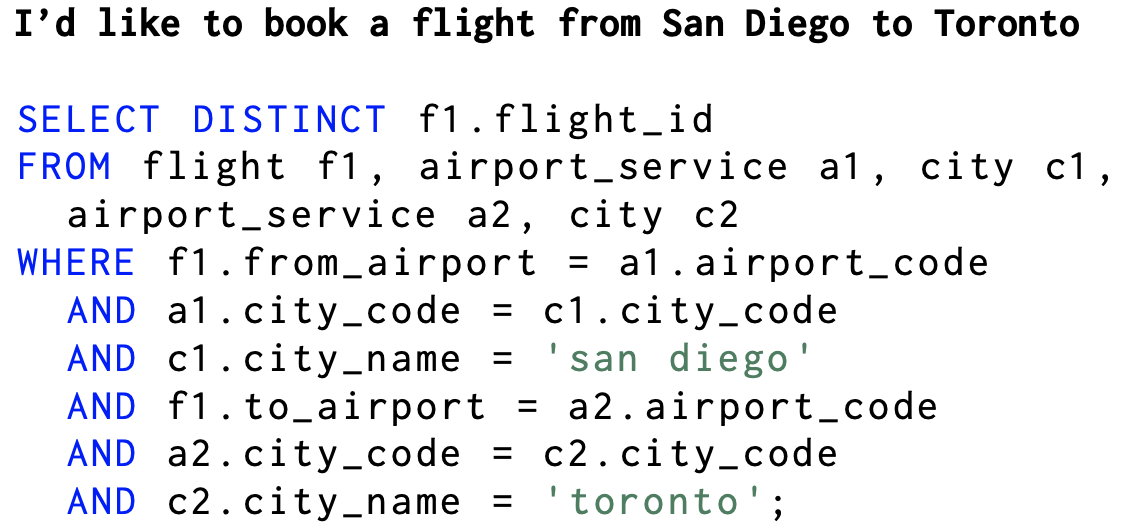}
    \caption{Sample SQL query for a flight reservation (credit:~\citet{iyer:2017:acl}).}
    \label{fig:sql}
\end{figure}

\begin{figure}
    \includegraphics[width=\columnwidth]{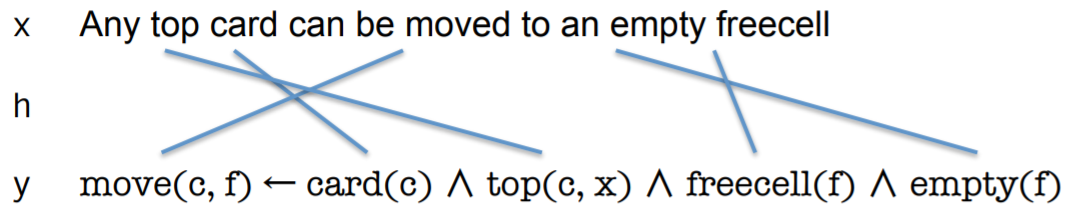}
    \caption{A natural language sentence (denoted x), a logical formula corresponding to the
    sentence (denoted y), and a lexical alignment between the two (denoted h) (credit:~\citet{goldwasser:2014:mljournal}).}
    \label{fig:freecell}
\end{figure}

Meaning representations are the end product of NL semantic parsing. The output logical formula, or meaning representation, grounds the semantics of the natural language input in the target machine-understandable domain. A simple example of logical forms is a database query. Popular datasets explored in semantic parsing include Geo880, Jobs640, ATIS, and SQL itself \citep{iyer:2017:acl,yu:2018:emnlp,dasigi:2019:acl}. An example of a SQL query is shown in Figure~\ref{fig:sql}. Another class of semantic parsing outputs is NL instructions to actions, such as the model in Figure~\ref{fig:freecell} for learning Freecell card game moves. Note that in these cases, the semantic parser is specific to a particular domain, and may not generalize to out-of-domain inquiries or differing data structures.

NL semantic parsing efforts have also translated natural language input semantics into lambda calculus expressions. The expressive power of lambda calculus allows the incorporation of constants, logical connectors, quantifications, and lambda expressions to represent functions \citep{zettlemoyer:2005:auai}.

\begin{tabbing}
\quad\=Sentence: \quad\quad\=what countries border France\\
\>Logical Form:\>$\lambda x.country(x) \wedge borders(x,France)$
\end{tabbing}

Meaning representations can also take on graph-based forms. Graph-based formalisms are advantageous because they are easier for humans to read and they are abstracted from the syntax of the input natural language sentences. Popular graphical representations are the Abstract Meaning Representation \cite{banarescu-etal-2013-abstract}, Semantic Dependency Parsing \cite{oepen-etal-2014-semeval,che-etal-2016-semeval}, Universal Conceptual Cognitive Annotation \cite{abend-rappoport-2013-universal}, Dependency-based Compositional Semantics \cite{liang:2011:acl}, and CCGBank \cite{hockenmaier:2007:computationallinguistics}. 

While most general purpose programming languages are not necessarily abstract in their representations of intention (source code often involves details about variables, memory allocation, class structure, etc.), code can serve as one of the more fundamental machine-understandable logical forms. To lift semantics from the code's syntax, recent works have gone into developing and using meaning representations of code, such as the abstract syntax tree (AST) \citep{alon:2019:acm,raychev:2015:popl,raychev:2016:oopsla,quirk-etal-2015-language,alon:2018:iclr,allamanis:2018:iclr}, sketching \citep{10.5555/1714168}, simplified parse tree (SPT) \citep{luan:2019:acm}, contextual flow graph (XFG) \citep{bennun:2018:nips}, context-aware semantic structure (CASS) \citep{ye2020misim}, and program-derived semantics graph (PSG) \citep{iyer2020software}. While none of these representation structures are identical, they somewhat share the common goal of being a vehicle of machine-understandable meaning of the code, abstracted to some extent from the syntax of code. We discuss the application of some of these structures in Section \ref{sec:program-synthesis}.

\section{Program Synthesis / Code Generation}
\label{sec:program-synthesis}
In this section, we review a subset of existing efforts for program synthesis and code generation. Program synthesis is the task of generating software from some high-level program specification \cite{zohar:1980:acm}. The purpose of this section is to highlight parallels between this process of using a high-level program specification to generate code, and semantic parsing's translation of natural language into meaning representations. Insights from the fields of program synthesis and machine programming can guide semantic parsing efforts, and vice versa.

Code has many different dimensions. Despite what could be construed as a strict syntax for compile-ability, source code can be reasoned about on many different levels: semantics, runtime, memory footprint, and compiler details, to name a few. As a result, design and selection of a code representation structure, and the subsequent reasoning about said structure is an interesting and nuanced task.

Many code recommendation systems use code similarity to retrieve different implementations of the same intention. Semantic code representations can provide a foundation on which to extract code similarity. Facebook et al.’s Aroma system for code recommendation \citep{luan:2019:acm} represents a program by a simplified parse tree (SPT): the parse tree is defined as a recursive structure of keyword and non-keyword tokens, agnostic to implementation programming language. Intel et al.’s MISIM system for code similarity \citep{ye2020misim} defines the context-aware semantic structure (CASS), which leverages the structural advantages of the SPT, but also carries the advantage of incorporating customizable parameters for a wider variety of code context scenarios. The CASS enables configuration at different levels of granularity, allowing it to be more flexible for general, rather than domain-specific, usage. \citet{iyer2020software} present the program-derived semantics graph (PSG) to capture semantics of code at multiple levels of abstraction. Additionally, it offers one layer of concrete syntactic abstraction that is programming language specific, which allows it to be converted into source code.

Program synthesis has also come in the flavor of machine translation \cite{wong:2006:naacl,quirk-etal-2015-language,chen2018treetotree}, pseudocode-to-code transformation \cite{kulal:2019:neurips}, and inductive program synthesis by input-output examples \cite{balog:2017:iclr,gulwani:2017:aplas,pmlr-v28-menon13}. While machine translation found its infancy in translation between human languages, a series of works has gone into translating from natural language to programming languages. \citet{wong:2006:naacl} integrate machine translation techniques with semantic parsing to create a syntax-based WASP: word alignment-based semantic parsing. \citet{quirk-etal-2015-language} further combine machine translation to extract ASTs consistent with a constructed grammar for If-This-Then-That (IFTTT) recipes. \citet{chen2018treetotree} tackle program translation by employing tree-to-tree deep neural networks to migrate code from one language into an ecosystem built in a different language. \citet{kulal:2019:neurips} perform the task of mapping line-to-line pseudocode to code for program synthesis. Inductive program synthesis allows us to produce a program consistent with a set of input-output examples \citep{balog:2017:iclr,gulwani:2017:aplas,pmlr-v28-menon13,pu:2018:pmlr,feser:2015:sigplan,lieberman:2001:book,perelman:2014:sigplan,gulwani:2011:sigplan}. In these approaches, the program synthesis problem can be framed as the following: first, encode input-output examples as constraints; then, search over the space of possible programs with a constraint solver. A key challenge in this technique is scalability. \citet{pu:2018:pmlr} demonstrate that inductive program synthesis can be made a scalable task by having a model iteratively select a small subset of representative examples, then train on each small subset at a time. Another aspect to consider is evaluation. A fitness function determines how well a given candidate program satisfies some set of requirements. Given these requirements as a set of input-output examples, such as in inductive program synthesis, the fitness function can be learned \cite{mandal:2021:mlsys}, resulting in more efficient program synthesis because it is biased to search likely programs.

The task of program synthesis tends to be more tractable if the output programming language is less complex and thus the output space is more compact. Declarative programming languages such as SQL \cite{chamberlin:1974:acm}, Halide \cite{ragankelley:2013:acm}, and GraphIT \cite{zhang:2018:acm} express the intention of a computation without detailing its execution. For this reason, these types of languages are also called \emph{intentional programming languages} in the field of machine programming. As a function of the syntax and semantics being closely related, programs written in intentional programming languages tend to be easier to reason about in a semantic space. A byproduct of this phenomenon is that intentional programming languages can also be easier to optimize. Neo \cite{marcus:2019:vldb} and Bao \cite{marcus2020bao} are learned neural models that optimize SQL queries. Verified lifting \cite{kamil:2016:pldi} also demonstrates an ability to optimize Halide code. 


As we discuss the evolution of semantic parsing in the next few sections, we identify parallels between these techniques used for program synthesis techniques and the techniques used in semantic parsing.
\vspace{-2mm}

\section{Evolution of Semantic Parsing}
\label{sec:evolution}
This section will review the evolution of NL semantic parsing techniques from early hand crafted rules in the 1970s to the rise of learned grammars via statistical learning methods in the late 1990s and early 2000s, to modern neural semantic parsing efforts. 

\subsection{Rules-based Techniques for Semantic Parsing}

Many early approaches to NL semantic parsing used hand-crafted syntactic pattern and phrase matching to parse the meaning from a given natural language input \citep{johnson_1984,woods:1973:ncce}. These rules were then augmented by grammars that incorporate knowledge of semantic categories \citep{waltz:1978:acm,lockemann-thompson-1969-rapidly,hendrix:1978:acm} so that similar words can be generalized to sets of defined categories and then constrain the output space of the logical forms. However, as a function of being based off of pattern-matching rules, these systems are brittle, and as a function of relying on defined semantic categories, these grammar-constrained systems are domain-specific.

\subsection{Grammar and Rule Induction for Semantic Parsing}

Statistical methods for the induction of NL semantic parsing grammars rose in the late 1990s and early 2000s. \citet{zelle:1996:aaai} present a system that learns rules for language parsing instead of relying on hand-crafting them. The system employs inductive logic programming methods to learn shift-reduce parsers that map sentences into database queries. \citet{kate:2005:ncai} also induce transformation rules that map natural language to logical form. By learning and recursively applying these rules, Kate et al.’s system maps natural language inputs (or their corresponding syntactic parse trees) to a parse tree in the target representation language. 

Other supervised learning approaches have induced grammars to represent the underlying semantics of a knowledge base. In the literature, CCGs have been a popular and powerful tool to capture textual semantics by mapping sentences to their logical forms \cite{artzi:2013:tacl,zettlemoyer:2005:auai,zettlemoyer-collins-2007-online,artzi-zettlemoyer-2011-bootstrapping}. 
\citet{artzi_zettlemoyer_2015} outline a CCG-based NL semantic parsing framework that maps sentences to lambda-calculus representations of their meaning. They induce a grammar and estimate the parameters of the compositional and non-compositional semantics of the space \cite{artzi-etal-2014-semantic}. Artzi’s approach uses corpus-level statistics at each lexicon update during learning to prune the induced CCG lexicons. Performance improvements in this system indicate an advantage to maintaining compact CCG lexicons. However, these model are typically unable to learn entities not presented to the model during training. \citep{artzi-etal-2014-learning,zettlemoyer-collins-2007-online}. 

A challenge in semantic parsing has been how to scale a semantic parsing model for different domains. \citet{kwiatkowski-etal-2013-scaling} present a learned ontology matching model that adapts the output logical form of the semantic parser for each target domain. The parser uses a probabilistic CCG to construct a domain-independent meaning representation from the NL input, then uses the ontology matching model to transform that representation into the target domain. 

\subsection{Neural Models for NL Semantic Parsing}
\begin{figure}
\includegraphics[width=\linewidth]{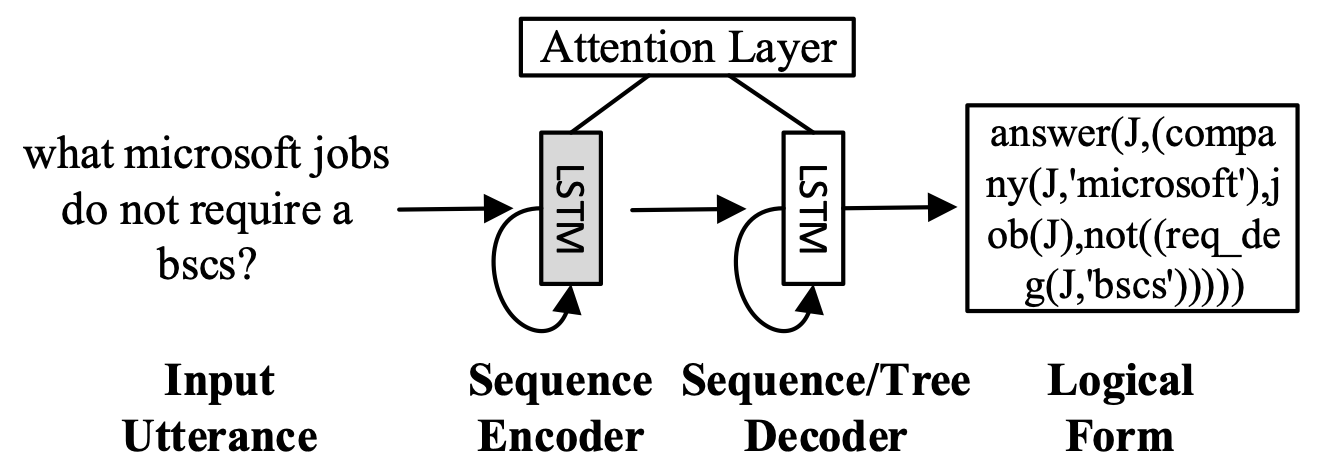}
\caption{Encoder-decoder model for mapping NL to logical forms (Credit:~\citet{dong2016language}).}\label{fig:encoderdecoder}
\end{figure}

The modern rise in neural models has been marked by the increase in papers exploring encoder-decoder sequence-to-sequence (seq2seq) frameworks for semantic parsing. These encoders and decoders are often built with recurrent architectures such as LSTM cells (see Figure~\ref{fig:encoderdecoder}) 
in an effort to better capture the dynamic and sequential nature of both natural and programming languages. Neural models can enable systems to automatically induce grammars, templates, and features instead of relying on humans to manually annotate them. This provides the model with flexibility to generalize across languages and representations \cite{krishnamurthy:2017:empiricalmethodsnl}. These encoder-decoder models can bear a resemblance to machine translation systems \cite{wong:2006:naacl} except that NL semantic parsing systems produce a structured logical form, while machine translation systems traditionally translate a natural language input into another unstructured natural language output \cite{andreas:2013:acl}. The formal output structure of semantic parsing enables neural decoders to lean on the logic of the induced grammars, while retaining the simplicity of neural translation without latent syntactic variables\cite{krishnamurthy:2017:empiricalmethodsnl,yin-neubig-2017-syntactic,rabinovich:2017:acllong}. Thus, this type of modeling can generalize to different environments while reducing demand for custom-annotated datasets. It should be noted, however, that these neural models tend to have a weakened ability to leverage the logical compositionality of traditional rules-based approaches.

\citet{dong2016language} present a NL semantic parsing model that employs an attention-enhanced encoder-decoder. In a nod to the hierarchical structure of their system's logical forms, they design a top-down tree decoder to more faithfully decode well-formed meaning representations. Decoding of each token is augmented by: \emph{(i)} soft alignments from attention computed by the encoder and \emph{(ii)} parent-feeding connections to provide contextual information from prior time steps. The resulting model has competitive performance and increased flexibility without need for any human-engineered features, making it more scalable to different domains. This work marks the advent of growing attention to neural models for semantic parsing.

A more in-depth discussion of modern techniques in neural semantic parsing models is presented in Section~\ref{sec:modern}, where we focus on semantic parsing models for code generation. 

\section{Supervision in Semantic Parsing}
\label{sec:supervision}
Parallel to the evolution of NL semantic parsing techniques is the evolution of supervision for semantic parsing. Fully-supervised learning customarily mandates a corpus of paired natural language utterances and their corresponding logical forms. However, this can be expensive to generate and difficult to scale. 

\paragraph{Supervision by Denotation} Supervision by denotation is proposed as an alternative, more scalable approach to NL semantic parsing. Instead of directly evaluating the success of a semantic parser by comparing the output to a gold standard “correct” logical form as done in fully-supervised learning, NL semantic parsing can be driven by the response of the context or environment. \citet{clarke:2010:conll} present a learning algorithm that relies only on a binary feedback signal for validation. The resulting parser is competitive with fully supervised systems in the Geoquery domain, demonstrating the feasibility of supervision by denotation. \citet{liang:2011:acl} also elide the need for labeled logical forms by inducing a NL semantic parsing model from question-answer pairs. They introduce a new semantic representation, the \emph{dependency-based compositional semantics (DCS)} to represent formal semantics as a tree. The model maps a NL utterance to DCS trees, aligns predicates within each tree to the span in the utterance that triggered it, and beam search for the best candidates.

However, at least two primary challenges arise with supervision by denotation. \emph{(1)}	The space of potential denotations may be large. This can increase computational tractability challenges in the search problem for intermediate latent representations. \emph{(2)} A weaker response signal has the potential for spurious logical forms: semantic parsing outputs that happen to produce the correct response, but do not actually carry the correct semantics. Modifications to supervision by denotation have been proposed to address these challenges. \citet{goldwasser:2014:mljournal} present a method of learning from natural language instructions and a binary feedback signal to address these issues.
By capturing alignments between NL instruction textual fragments and meaning representation logical fragments, the model can more closely interpret NL instructions and hopefully reduce the potential for spurious forms. 
\citet{pasupat-liang-2016-inferring} propose a system that performs dynamic programming on denotations to address the issue of exploding latent search space and spurious results. By observing that the space of possible denotations grows more slowly than the space of logical forms, this dynamic programming approach can find all consistent logical forms up to some bounded size. Then, because spurious logical forms will generally give a wrong answer once the context changes, they generate a slightly modified context and examine whether the logical form produces the correct denotation in this alternate context. If it does not, this spurious logical form can be discarded. The combination of limited search space and pruning of spurious forms allows them to use denotation as a stronger supervision signal. 

\paragraph{Weakly Supervised Learning} \citet{artzi-zettlemoyer-2011-bootstrapping} use conversational feedback from un-annotated conversation logs as a form of weak supervision to induce a NL semantic parser. In this approach, Artzi and Zettlemoyer model the meaning of user statements in the conversational log with latent variables, and define loss as a rough measure of how well a candidate meaning representation matches the expectations about the conversation context and dialog domain. 
\citet{artzi:2013:tacl} demonstrate the viability of weak supervision for mapping natural language instructions to actions in the robotics domain. They run the output of a parsing component through an execution component and observe the result relative to a binary-feedback validation function in the environment.~\footnote{We discuss other emerging works of weak supervision, such as Snorkel~\cite{ratner:2019:vldb}, in Section~\ref{sec:future}.}


\paragraph{Self-supervised Learning}
\citet{poon:2009:emnlp} present a self-supervised system for NL semantic parsing by recursively clustering semantically equivalent logical forms to abstract away syntactical implementation details. This method, however, does not ground the induced clusters in an ontology. \citet{poon-2013-grounded} subsequently proposes a \emph{grounded} unsupervised learning approach that takes a set of natural language questions and a database, learns a probabilistic semantic grammar over it, then constrains the search space using the database schema. For a given input sentence, this system annotates the sentence’s dependency tree with latent semantic states derived from the database. The dependency-based logical form is additionally augmented with a state space to account for semantic relationships not represented explicitly in the dependency tree. This semantically-annotated tree can then be deterministically converted into the logical form, without ever needing pre-labeled data. 
\citet{goldwasser-etal-2011-confidence} suggest a confidence-driven self-supervised protocol for NL semantic parsing. The model compensates for the lack of training data by employing a self-training protocol driven by confidence estimation to iteratively re-train on high-confidence predicted datapoints.

Open vocabulary NL semantic parsing replaces a formal knowledge base with a probabilistic base learned from a text corpus. In this framework, language is mapped onto queries with predicates (parts of the structure that make claims about the actions or characteristics of the subject) derived directly from the domain text instead of provided through a grammar. The model must also learn the execution models for these induced predicates \cite{krishnamurthy:2015:tacl,lewis:2013:tacl}. \citet{gardner:2017:aaai} address the challenge of generalizing NL semantic parsing models to out-of-domain and open-vocabulary applications. Instead of mapping language to a single knowledge base query like traditional NL semantic parsers, this approach maps language to a weighted combination of queries plus a distributional component to represent a much broader class of concepts. This approach can expand the applicability of semantic parsing techniques to more complex domains.

\section{Modern Advances in Neural Semantic Parsing for Code Generation}
Recent semantic parsing efforts have largely centered around neural models, especially encoder-decoder networks. In this section, we discuss modern advances in neural models for semantic parsing for code generation.
\label{sec:modern}
\subsection{Context and Variable Copying}
Information to generate a NL semantic parse is not always encapsulated entirely in a single input sentence. Surrounding context can provide valuable information for disambiguation in NL semantic parsing \cite{artzi-zettlemoyer-2011-bootstrapping}. Context can be fed into the encoder to incorporate context to encoder-decoder models.

\citet{iyer-etal-2018-mapping} develop an architecture that leverages programmatic context to write contextually-relevant code. This encoder-decoder system feeds the encoder the input utterance along with featurized representations of environment identifiers, such as type and name of each variable and method. The decoding process attends first to the natural language, then to the variables and methods of the environment. Performing this attention in two steps can help the model match words in the natural language to the environment identifier representations. Additionally, a supervised copy mechanism \cite{gu:2016:acl} based on attention weights is presented to copy an environment token to the output. This allows the model to copy in variable names not seen in the training data or in the input utterance.

\subsection{Grammar in Neural Semantic Parsing Models}
Grammar also can guide and constrain the decoding process of neural encoder-decoder models. \citet{xiao-etal-2016-sequence} propose a sequence-based approach to generate a knowledge base query: they sequentialize the target logical forms, then demonstrate the advantage of incorporating grammatical constraints into the derivation sequence. Further, when querying information from semi-structured tables, \citet{krishnamurthy:2017:empiricalmethodsnl} demonstrate the benefits of enforcing type constraints during query generation. They show that including explicit entity linking during encoding can bolster accuracy because it ensures the generation of well-typed logical forms.

As discussed in Section~\ref{sec:program-synthesis}, code is dynamic in its representation and semantics, making grammar especially interesting for generating source code. Unlike prior works which focus on domain-specific languages, which may have limited scope and complexity, \citet{yin-neubig-2017-syntactic} and \citet{ling2016latent} present methods for generating high-level general-purpose programming language code such as Python. \citet{ling2016latent} present a data-driven sequence-to-sequence code generation method that implements multiple predictors: a generative model for code generation, and multiple pointer models to copy keywords from the input. Building off of \citet{ling2016latent}, \citet{yin-neubig-2017-syntactic} point out that general purpose languages are generally more complex in schema and syntax than other domain-specific semantic parsing challenges. Therefore, code generation models could benefit from a grammar that explicitly captures the target programming language syntax as prior knowledge. \citet{yin-neubig-2017-syntactic} use abstract syntax trees (ASTs) to constrain the search space by exclusively allowing well-formed outputs: the model generates a series of actions that produce a valid AST. The actions themselves are derived from the underlying syntax of the programming language. This approach demonstrates improved Python code generation, but Yin and Neubig point out a decline in performance as the AST becomes more complex, noting that there may be space for improvement in scalability.

Code idioms are bits of code that occur frequently across a code repository and tend to have a specific semantic purpose. They can be used to enhance grammars in semantic parsing. \citet{iyer2019learning} apply idiom-based decoding to reduce the burden on training data in grammar-constrained semantic parsing systems. The top-K most frequently seen idioms in the training set are used to collapse the grammar for decoding. This compact set of rules helps supervise the learning of the semantic parser. However, there is a limit to how much idioms help with performance. The paper notes that adding too many idioms may cause the model to over-fit and become less generalizable. \citet{shin2019program} mine code idioms with the objective of helping models reason about high-level and low-level reasoning at the same time. The mined code idioms are incorporated into the grammar as new named operators, allowing the model to emit entire idioms at once. Incorporating the code idioms into the vocabulary of the neural synthesizer enables it to generate high-level and low-level program constructs interchangeably at each step, which is more like how human programmers write code.


\subsection{Sketching}
\citet{10.1145/1168919.1168907} present the concept of sketching to generate programs from a high-level description. Through sketching, developers communicate insight through a partial program, i.e. the sketch, that describes the high-level structure of an implementation. Then, synthesis procedures can generate the full implementation of this sketch in programming language. 

\citet{dong-lapata-2018-coarse} present a semantic parsing decoder that operates at varying levels of abstraction for encoder-decoder semantic parsing. By operating as a representation of the basic meaning, the sketch can provide a basic idea of the meaning of the utterance while abstracting away from the syntactic details of the actual implementation. The decoder presented in this paper operates in two stages: (1) generating a coarse sketch of the meaning representation that omits low-level details, and (2) filling in those missing details of arguments and variable names. This coarse-to-fine decoding has the advantage of disentangling high and low-level semantic information so that the program can be modeled at different granularities. An attention-enhanced RNN serves as the fine meaning decoder, using the coarse sketch to constrain the decoding output. This approach does not utilize syntax or grammatical information because the sketch naturally constrains the output space. 

\citet{nye2019learning} present a method to dynamically integrate natural language instructions and examples to infer program sketches. Previous systems have mostly used static, hand-designed intermediate sketch grammars, which tend to be rigid in how much a system relies on pattern recognition vs. symbolic search. When given an easy or familiar task, \citet{nye2019learning}'s neuro-symbolic synthesis system can rely on learned pattern recognition to produce a more complete output. When given a hard task, the system can produce a less complete sketch and spend more time filling in that sketch using search techniques. By learning how complete to make the sketch for different applications, this model can employ sketching in a more fitting and customized manner. 

\subsection{Conversational Semantic Parsing}
\label{sec:conversational}
A remaining challenge in neural semantic parsers is interpretability. \citet{dong2018confidence} outline a confidence modeling framework that characterizes three uncertainties: model uncertainty, data uncertainty, and input uncertainty. By computing these confidence metrics for a given prediction and using them as features in a regression model, they obtain a confidence score which can be used to identify which part of the input or model contributes to uncertain predictions. This uncertainty modeling allows models to better understand what it does and does not know. The ability to target weak parts of a model leads us to conversational programming approaches, where the model can query the user for missing or uncertain information.

When performing NL semantic parsing to extract some logical flow, conversational AI allows models to remedy incomplete or invalid requests. By engaging in a back-and-forth dialogue between the program and the human user, gaps and errors in logical forms can be filled and remedied, respectively. Additionally, clarifications can reduce the search space to promise a more accurate output \citep{artzi-zettlemoyer-2011-bootstrapping,chaurasia-mooney-2017-dialog,machines2020task-oriented}.

\section{Future Directions} 
\label{sec:future}
The task of connecting human communication with machine data and capabilities, as in the intention pillar of machine programming, has seen considerable headway over the past few decades. Moving forward in semantic parsing, we identify four areas of consideration for the next steps of semantic parsing for code generation:

\paragraph{Evaluation} Current methods of evaluation for semantic parsing correctness use strict matches to gold standards, denotation, hand-wavy heuristics, or some other imperfect form of measuring understanding of semantic meaning. Exact match evaluation may incorrectly penalize mismatching logical forms that actually produce the same semantic meaning. Supervision by denotation can incorrectly reward spurious logical forms. Hand-wavy heuristics such as Bleu score \citep{papineni-etal-2002-bleu}, which evaluates the "quality" of machine translation outputs via a modified n-gram precision metric, have been shown to not reliably capture semantic equivalence for natural language \citep{callison-burch-etal-2006-evaluating} or for code \citep{Tran_2019}. Efforts to resolve the weaknesses of each of these evaluation methods \cite{goldwasser:2014:mljournal,dasigi:2019:acl,pasupat-liang-2016-inferring,poon:2009:emnlp,poon-2013-grounded} have not been ineffective, but do not seem to fully address the key issue, which is that an evaluator for semantic parsing should have an understanding of semantic \emph{reasoning}. We propose that an important next step may be to research either or both of: \emph{(i)} an evaluator that can identify matching core semantics of different logical forms, and \emph{(ii)} a meaning representation that has one unique representation for any given semantic intention.

Evaluation of a model's actual understanding could be another open challenge in semantic parsing. Rules- and grammar-based systems \cite{artzi_zettlemoyer_2015,hendrix:1978:acm,johnson_1984,kate:2005:ncai,lockemann-thompson-1969-rapidly,waltz:1978:acm,zelle:1996:aaai} are generally more transparent and thus more interpretable than neural models because the rules on which they are built can be examined as the logic of the model. However, neural models tend to function more as "black boxes," where induced logic is mostly evaluated by examining the outputs. We recommend research moving forward to consider how to design this neural learning to better capture the complex nature of both natural language and source code in translating natural language into source code. Compositional generalization is one way to evaluate how well a model understands natural language. \citet{oren:2020:emnlp} investigate compositional generalization in semantic parsing. They find that while well-known and novel extensions to the seq2seq model improve generalization, performance on in-distribution data is still significantly higher than on out-of-distribution data. This suggests a need to make more drastic changes to the neural encoder-decoder approach in order to inject fundamental "understanding." Neural Module Networks (NMNs) can also provide interpretability, compositionality, and improved generalizability to semantic parsers \cite{gupta:2020:iclr,subramanian:2020:acl}. \citet{gupta:2020:iclr} investigate how multiple task-specialized NMNs can be composed together to perform complex reasoning over unstructured knowledge such as natural language text. However, the advanced interpretability of NMNs comes at the cost of restricted expressivity because each NMN needs to be individually defined. \citet{gupta:2020:iclr} reflect that this tradeoff suggests future further research to bridge these reasoning gaps.

\paragraph{Code Semantics Representation} We perceive that an ongoing challenge in semantic reasoning about language and code may be the lack of a unified code representation that can capture semantics and details enough to resolve ambiguity but remain agnostic of programming language. The graph-based meaning representations discussed in Section \ref{sec:meaningrep} are promising avenues to explore. They have shown to be effective in code retrieval, so we propose they be deployed in a semantic parsing framework to evaluate their viability for semantic parsing and code generation.

Big picture, developing this language-agnostic meaning representation is a promising direction to facilitate a full machine programming pipeline. If the meaning representation is transformable in all the directions between natural language, meaning representation, and code, this unified intermediate representation enables a sort of \emph{"programming by conversation."} Not only could software be generated from language, as discussed in this paper, it could also then be translated from source code into natural language. Then once in natural language form, intention could be iterated upon via an interactive conversational agent. The iterated representation could then be used to translate this refined intention back into software.

\paragraph{Scope} Most current semantic parsing models reduce the scope of the problem (and thus, often, output space) to make the problem more feasible. This can be done either by working with DSLs, using a selected subset of APIs, or by working only with code snippets of a limited length and complexity. However, actual functional programs tend to be longer and more complex than the code snippet outputs in current semantic parsing models. Generation of programs in a GPL \citep{ling2016latent} is generally a more complex task than in a DSL due to the wide variety of domains and the high-dimensional code reasoning space. Eventually, we want to be able to work dynamically with larger programs. Steps toward incorporating context to and using conversational AI in neural semantic parsing models have shown promising results, and should continue to be pursued. 

As the field moves into working with larger programs, existing datasets that match natural language utterances to code snippets will likely not be enough. To obtain more data for neural deep learning, we recommend using data generation models such as Snorkel \cite{ratner:2019:vldb}. Snorkel employs weak supervision to create training data without any need for hand-labeled data. We also suggest that new datasets be developed that account for longer chunks of code. This can be done by factoring in context variables and functions, or by mining larger chunks of code with periodic natural language annotations in the form of comments. 

\paragraph{User Interfacing} Future work could also consider the actual impact of code generation on programmer productivity. \citet{xu2021inide} conducted a study that found no impact on code correctness or generation efficiency when the subjects were asked to work in a programming environment IDE with a code generation plug-in. 
This begs the question of how to develop future models to be not just user-friendly, but user aiding. Additionally, user interfaces for code generation may not share the same vocabulary between trained programmers and non-technical developers. There may be room for natural language understanding research advances for models to intuit programming concepts from plain English, free of such jargon as \emph{“list,”} \emph{“graph,”} and \emph{“iterate.”}




\section{Broader Impact}
The objective of this paper is to compile information and synthesize ideas about the fields of semantic parsing for natural language and for code. Since the basis of this paper is primarily educational, we see its broader impacts to largely be positive.

Semantic parsing for code generation will likely grant more people a greater capacity to create programs. However, this enhanced capability also carries potential negative impacts. One possible misuse is the creation of malicious code. The models outlined in this paper could be trained to learn to write computer viruses, identify security breaches in other peoples' networks, or any of a number of dangerous behaviors. Aside from continuing to make this malicious code illegal, it may be worthwhile to embed safety features into the prior knowledge of these models, to discourage destructive code. On the flip side, models for code generation could also in the worst case generate programs with egregiously poor performance. If deployed, these programs will consume excessive computation and energy, thus negatively impacting the environment. One way to mitigate this harm is to encourage code reviews for generated code, just as code reviews exist now for human-created code. Another potential misuse of this technology that we would be concerned about is computer science students using this technology to complete their programming assignments, rather than learning how to write them themselves. This will be difficult to discourage from the model development side, but development and usage of a technology to identify coding style could help instructors detect whether students are writing their own code. Overall, researchers, as designers of these semantic parsing systems, should consider the safety and impact of their work on the many different realms of society. 

\bibliography{ms}


\begin{thebibliography}{94}


\ifx \showCODEN    \undefined \def \showCODEN     #1{\unskip}     \fi
\ifx \showDOI      \undefined \def \showDOI       #1{#1}\fi
\ifx \showISBNx    \undefined \def \showISBNx     #1{\unskip}     \fi
\ifx \showISBNxiii \undefined \def \showISBNxiii  #1{\unskip}     \fi
\ifx \showISSN     \undefined \def \showISSN      #1{\unskip}     \fi
\ifx \showLCCN     \undefined \def \showLCCN      #1{\unskip}     \fi
\ifx \shownote     \undefined \def \shownote      #1{#1}          \fi
\ifx \showarticletitle \undefined \def \showarticletitle #1{#1}   \fi
\ifx \showURL      \undefined \def \showURL       {\relax}        \fi
\providecommand\bibfield[2]{#2}
\providecommand\bibinfo[2]{#2}
\providecommand\natexlab[1]{#1}
\providecommand\showeprint[2][]{arXiv:#2}

\bibitem[\protect\citeauthoryear{Abend and Rappoport}{Abend and
  Rappoport}{2013}]%
        {abend-rappoport-2013-universal}
\bibfield{author}{\bibinfo{person}{Omri Abend} {and} \bibinfo{person}{Ari
  Rappoport}.} \bibinfo{year}{2013}\natexlab{}.
\newblock \showarticletitle{{U}niversal {C}onceptual {C}ognitive {A}nnotation
  ({UCCA})}. In \bibinfo{booktitle}{\emph{Proceedings of the 51st Annual
  Meeting of the Association for Computational Linguistics (Volume 1: Long
  Papers)}}. \bibinfo{publisher}{Association for Computational Linguistics},
  \bibinfo{address}{Sofia, Bulgaria}, \bibinfo{pages}{228--238}.
\newblock
\urldef\tempurl%
\url{https://www.aclweb.org/anthology/P13-1023}
\showURL{%
\tempurl}


\bibitem[\protect\citeauthoryear{Allamanis, Brockschmidt, and
  Khademi}{Allamanis et~al\mbox{.}}{2018}]%
        {allamanis:2018:iclr}
\bibfield{author}{\bibinfo{person}{Miltiadis Allamanis}, \bibinfo{person}{Marc
  Brockschmidt}, {and} \bibinfo{person}{Mahmoud Khademi}.}
  \bibinfo{year}{2018}\natexlab{}.
\newblock \showarticletitle{Learning to Represent Programs with Graphs}. In
  \bibinfo{booktitle}{\emph{International Conference on Learning
  Representations}}.
\newblock


\bibitem[\protect\citeauthoryear{Alon, Levy, and Yahav}{Alon
  et~al\mbox{.}}{2019a}]%
        {alon:2018:iclr}
\bibfield{author}{\bibinfo{person}{Uri Alon}, \bibinfo{person}{Omer Levy},
  {and} \bibinfo{person}{Eran Yahav}.} \bibinfo{year}{2019}\natexlab{a}.
\newblock \showarticletitle{code2seq: Generating Sequences from Structured
  Representations of Code}. In \bibinfo{booktitle}{\emph{International
  Conference on Learning Representations}}.
\newblock


\bibitem[\protect\citeauthoryear{Alon, Zilberstein, Levy, and Yahav}{Alon
  et~al\mbox{.}}{2019b}]%
        {alon:2019:acm}
\bibfield{author}{\bibinfo{person}{Uri Alon}, \bibinfo{person}{Meital
  Zilberstein}, \bibinfo{person}{Omer Levy}, {and} \bibinfo{person}{Eran
  Yahav}.} \bibinfo{year}{2019}\natexlab{b}.
\newblock \showarticletitle{Code2vec: Learning Distributed Representations of
  Code}.
\newblock \bibinfo{journal}{\emph{Proc. ACM Program. Lang.}}
  \bibinfo{volume}{3}, \bibinfo{number}{POPL}, Article \bibinfo{articleno}{40}
  (\bibinfo{date}{Jan.} \bibinfo{year}{2019}), \bibinfo{numpages}{29}~pages.
\newblock
\urldef\tempurl%
\url{https://doi.org/10.1145/3290353}
\showDOI{\tempurl}


\bibitem[\protect\citeauthoryear{Andreas, Vlachos, and Clark}{Andreas
  et~al\mbox{.}}{2013}]%
        {andreas:2013:acl}
\bibfield{author}{\bibinfo{person}{Jacob Andreas}, \bibinfo{person}{Andreas
  Vlachos}, {and} \bibinfo{person}{Stephen Clark}.}
  \bibinfo{year}{2013}\natexlab{}.
\newblock \showarticletitle{Semantic Parsing as Machine Translation}. In
  \bibinfo{booktitle}{\emph{Proceedings of the 51st Annual Meeting of the
  Association for Computational Linguistics (Volume 2: Short Papers)}}.
  \bibinfo{publisher}{Association for Computational Linguistics},
  \bibinfo{address}{Sofia, Bulgaria}, \bibinfo{pages}{47--52}.
\newblock
\urldef\tempurl%
\url{https://www.aclweb.org/anthology/P13-2009}
\showURL{%
\tempurl}


\bibitem[\protect\citeauthoryear{Artzi, Das, and Petrov}{Artzi
  et~al\mbox{.}}{2014a}]%
        {artzi-etal-2014-learning}
\bibfield{author}{\bibinfo{person}{Yoav Artzi}, \bibinfo{person}{Dipanjan Das},
  {and} \bibinfo{person}{Slav Petrov}.} \bibinfo{year}{2014}\natexlab{a}.
\newblock \showarticletitle{Learning Compact Lexicons for {CCG} Semantic
  Parsing}. In \bibinfo{booktitle}{\emph{Proceedings of the 2014 Conference on
  Empirical Methods in Natural Language Processing ({EMNLP})}}.
  \bibinfo{publisher}{Association for Computational Linguistics},
  \bibinfo{address}{Doha, Qatar}, \bibinfo{pages}{1273--1283}.
\newblock
\urldef\tempurl%
\url{https://doi.org/10.3115/v1/D14-1134}
\showDOI{\tempurl}


\bibitem[\protect\citeauthoryear{Artzi, Fitzgerald, and Zettlemoyer}{Artzi
  et~al\mbox{.}}{2014b}]%
        {artzi-etal-2014-semantic}
\bibfield{author}{\bibinfo{person}{Yoav Artzi}, \bibinfo{person}{Nicholas
  Fitzgerald}, {and} \bibinfo{person}{Luke Zettlemoyer}.}
  \bibinfo{year}{2014}\natexlab{b}.
\newblock \showarticletitle{Semantic Parsing with {C}ombinatory {C}ategorial
  {G}rammars}. In \bibinfo{booktitle}{\emph{Proceedings of the 2014 Conference
  on Empirical Methods in Natural Language Processing: Tutorial Abstracts}}.
  \bibinfo{publisher}{Association for Computational Linguistics},
  \bibinfo{address}{Doha, Qatar}.
\newblock
\urldef\tempurl%
\url{https://www.aclweb.org/anthology/D14-2003}
\showURL{%
\tempurl}


\bibitem[\protect\citeauthoryear{Artzi and Zettlemoyer}{Artzi and
  Zettlemoyer}{2011}]%
        {artzi-zettlemoyer-2011-bootstrapping}
\bibfield{author}{\bibinfo{person}{Yoav Artzi} {and} \bibinfo{person}{Luke
  Zettlemoyer}.} \bibinfo{year}{2011}\natexlab{}.
\newblock \showarticletitle{Bootstrapping Semantic Parsers from Conversations}.
  In \bibinfo{booktitle}{\emph{Proceedings of the 2011 Conference on Empirical
  Methods in Natural Language Processing}}. \bibinfo{publisher}{Association for
  Computational Linguistics}, \bibinfo{address}{Edinburgh, Scotland, UK.},
  \bibinfo{pages}{421--432}.
\newblock
\urldef\tempurl%
\url{https://www.aclweb.org/anthology/D11-1039}
\showURL{%
\tempurl}


\bibitem[\protect\citeauthoryear{Artzi and Zettlemoyer}{Artzi and
  Zettlemoyer}{2013}]%
        {artzi:2013:tacl}
\bibfield{author}{\bibinfo{person}{Yoav Artzi} {and} \bibinfo{person}{Luke
  Zettlemoyer}.} \bibinfo{year}{2013}\natexlab{}.
\newblock \showarticletitle{Weakly Supervised Learning of Semantic Parsers for
  Mapping Instructions to Actions}.
\newblock \bibinfo{journal}{\emph{Transactions of the Association for
  Computational Linguistics}}  \bibinfo{volume}{1} (\bibinfo{year}{2013}),
  \bibinfo{pages}{49--62}.
\newblock
\urldef\tempurl%
\url{https://doi.org/10.1162/tacl_a_00209}
\showDOI{\tempurl}


\bibitem[\protect\citeauthoryear{Artzi and Zettlemoyer}{Artzi and
  Zettlemoyer}{2015}]%
        {artzi_zettlemoyer_2015}
\bibfield{author}{\bibinfo{person}{Yoav Artzi} {and} \bibinfo{person}{Luke~S.
  Zettlemoyer}.} \bibinfo{year}{2015}\natexlab{}.
\newblock \emph{\bibinfo{title}{Situated understanding and learning of natural
  language}}.
\newblock \bibinfo{thesistype}{Ph.D. Dissertation}.
\newblock


\bibitem[\protect\citeauthoryear{Balog, Gaunt, Brockschmidt, Nowozin, and
  Tarlow}{Balog et~al\mbox{.}}{2017}]%
        {balog:2017:iclr}
\bibfield{author}{\bibinfo{person}{Matej Balog}, \bibinfo{person}{Alexander~L.
  Gaunt}, \bibinfo{person}{Marc Brockschmidt}, \bibinfo{person}{Sebastian
  Nowozin}, {and} \bibinfo{person}{Daniel Tarlow}.}
  \bibinfo{year}{2017}\natexlab{}.
\newblock \showarticletitle{DeepCoder: Learning to Write Programs}. In
  \bibinfo{booktitle}{\emph{5th International Conference on Learning
  Representations, {ICLR} 2017, Toulon, France, April 24-26, 2017, Conference
  Track Proceedings}}.
\newblock


\bibitem[\protect\citeauthoryear{Banarescu, Bonial, Cai, Georgescu, Griffitt,
  Hermjakob, Knight, Koehn, Palmer, and Schneider}{Banarescu
  et~al\mbox{.}}{2013}]%
        {banarescu-etal-2013-abstract}
\bibfield{author}{\bibinfo{person}{Laura Banarescu}, \bibinfo{person}{Claire
  Bonial}, \bibinfo{person}{Shu Cai}, \bibinfo{person}{Madalina Georgescu},
  \bibinfo{person}{Kira Griffitt}, \bibinfo{person}{Ulf Hermjakob},
  \bibinfo{person}{Kevin Knight}, \bibinfo{person}{Philipp Koehn},
  \bibinfo{person}{Martha Palmer}, {and} \bibinfo{person}{Nathan Schneider}.}
  \bibinfo{year}{2013}\natexlab{}.
\newblock \showarticletitle{{A}bstract {M}eaning {R}epresentation for
  Sembanking}. In \bibinfo{booktitle}{\emph{Proceedings of the 7th Linguistic
  Annotation Workshop and Interoperability with Discourse}}.
  \bibinfo{publisher}{Association for Computational Linguistics},
  \bibinfo{address}{Sofia, Bulgaria}, \bibinfo{pages}{178--186}.
\newblock
\urldef\tempurl%
\url{https://www.aclweb.org/anthology/W13-2322}
\showURL{%
\tempurl}


\bibitem[\protect\citeauthoryear{Ben-Nun, Jakobovits, and Hoefler}{Ben-Nun
  et~al\mbox{.}}{2018}]%
        {bennun:2018:nips}
\bibfield{author}{\bibinfo{person}{Tal Ben-Nun},
  \bibinfo{person}{Alice~Shoshana Jakobovits}, {and} \bibinfo{person}{Torsten
  Hoefler}.} \bibinfo{year}{2018}\natexlab{}.
\newblock \showarticletitle{Neural Code Comprehension: A Learnable
  Representation of Code Semantics}. In \bibinfo{booktitle}{\emph{Proceedings
  of the 32nd International Conference on Neural Information Processing
  Systems}} (Montr\'{e}al, Canada) \emph{(\bibinfo{series}{NIPS'18})}.
  \bibinfo{publisher}{Curran Associates Inc.}, \bibinfo{address}{Red Hook, NY,
  USA}, \bibinfo{pages}{3589–3601}.
\newblock


\bibitem[\protect\citeauthoryear{Callison-Burch, Osborne, and
  Koehn}{Callison-Burch et~al\mbox{.}}{2006}]%
        {callison-burch-etal-2006-evaluating}
\bibfield{author}{\bibinfo{person}{Chris Callison-Burch},
  \bibinfo{person}{Miles Osborne}, {and} \bibinfo{person}{Philipp Koehn}.}
  \bibinfo{year}{2006}\natexlab{}.
\newblock \showarticletitle{Re-evaluating the Role of {B}leu in Machine
  Translation Research}. In \bibinfo{booktitle}{\emph{11th Conference of the
  {E}uropean Chapter of the Association for Computational Linguistics}}.
  \bibinfo{publisher}{Association for Computational Linguistics},
  \bibinfo{address}{Trento, Italy}.
\newblock
\urldef\tempurl%
\url{https://www.aclweb.org/anthology/E06-1032}
\showURL{%
\tempurl}


\bibitem[\protect\citeauthoryear{Chamberlin and Boyce}{Chamberlin and
  Boyce}{1974}]%
        {chamberlin:1974:acm}
\bibfield{author}{\bibinfo{person}{Donald~D. Chamberlin} {and}
  \bibinfo{person}{Raymond~F. Boyce}.} \bibinfo{year}{1974}\natexlab{}.
\newblock \showarticletitle{SEQUEL: A Structured English Query Language}. In
  \bibinfo{booktitle}{\emph{Proceedings of the 1974 ACM SIGFIDET (Now SIGMOD)
  Workshop on Data Description, Access and Control}} (Ann Arbor, Michigan)
  \emph{(\bibinfo{series}{SIGFIDET '74})}. \bibinfo{publisher}{Association for
  Computing Machinery}, \bibinfo{address}{New York, NY, USA},
  \bibinfo{pages}{249–264}.
\newblock
\showISBNx{9781450374156}
\urldef\tempurl%
\url{https://doi.org/10.1145/800296.811515}
\showDOI{\tempurl}


\bibitem[\protect\citeauthoryear{Chaurasia and Mooney}{Chaurasia and
  Mooney}{2017}]%
        {chaurasia-mooney-2017-dialog}
\bibfield{author}{\bibinfo{person}{Shobhit Chaurasia} {and}
  \bibinfo{person}{Raymond~J. Mooney}.} \bibinfo{year}{2017}\natexlab{}.
\newblock \showarticletitle{Dialog for Language to Code}. In
  \bibinfo{booktitle}{\emph{Proceedings of the Eighth International Joint
  Conference on Natural Language Processing (Volume 2: Short Papers)}}.
  \bibinfo{publisher}{Asian Federation of Natural Language Processing},
  \bibinfo{address}{Taipei, Taiwan}, \bibinfo{pages}{175--180}.
\newblock
\urldef\tempurl%
\url{https://www.aclweb.org/anthology/I17-2030}
\showURL{%
\tempurl}


\bibitem[\protect\citeauthoryear{Che, Shao, Liu, and Ding}{Che
  et~al\mbox{.}}{2016}]%
        {che-etal-2016-semeval}
\bibfield{author}{\bibinfo{person}{Wanxiang Che}, \bibinfo{person}{Yanqiu
  Shao}, \bibinfo{person}{Ting Liu}, {and} \bibinfo{person}{Yu Ding}.}
  \bibinfo{year}{2016}\natexlab{}.
\newblock \showarticletitle{{S}em{E}val-2016 Task 9: {C}hinese Semantic
  Dependency Parsing}. In \bibinfo{booktitle}{\emph{Proceedings of the 10th
  International Workshop on Semantic Evaluation ({S}em{E}val-2016)}}.
  \bibinfo{publisher}{Association for Computational Linguistics},
  \bibinfo{address}{San Diego, California}, \bibinfo{pages}{1074--1080}.
\newblock
\urldef\tempurl%
\url{https://doi.org/10.18653/v1/S16-1167}
\showDOI{\tempurl}


\bibitem[\protect\citeauthoryear{Chen, Liu, and Song}{Chen
  et~al\mbox{.}}{2018}]%
        {chen2018treetotree}
\bibfield{author}{\bibinfo{person}{Xinyun Chen}, \bibinfo{person}{Chang Liu},
  {and} \bibinfo{person}{Dawn Song}.} \bibinfo{year}{2018}\natexlab{}.
\newblock \showarticletitle{Tree-to-Tree Neural Networks for Program
  Translation}. In \bibinfo{booktitle}{\emph{Proceedings of the 32nd
  International Conference on Neural Information Processing Systems}}
  (Montr\'{e}al, Canada) \emph{(\bibinfo{series}{NIPS'18})}.
  \bibinfo{publisher}{Curran Associates Inc.}, \bibinfo{address}{Red Hook, NY,
  USA}, \bibinfo{pages}{2552–2562}.
\newblock


\bibitem[\protect\citeauthoryear{Clarke, Goldwasser, Chang, and Roth}{Clarke
  et~al\mbox{.}}{2010}]%
        {clarke:2010:conll}
\bibfield{author}{\bibinfo{person}{James Clarke}, \bibinfo{person}{Dan
  Goldwasser}, \bibinfo{person}{Ming-Wei Chang}, {and} \bibinfo{person}{Dan
  Roth}.} \bibinfo{year}{2010}\natexlab{}.
\newblock \showarticletitle{Driving Semantic Parsing from the World{'}s
  Response}. In \bibinfo{booktitle}{\emph{Proceedings of the Fourteenth
  Conference on Computational Natural Language Learning}}.
  \bibinfo{publisher}{Association for Computational Linguistics},
  \bibinfo{address}{Uppsala, Sweden}, \bibinfo{pages}{18--27}.
\newblock
\urldef\tempurl%
\url{https://www.aclweb.org/anthology/W10-2903}
\showURL{%
\tempurl}


\bibitem[\protect\citeauthoryear{Dasigi, Gardner, Murty, Zettlemoyer, and
  Hovy}{Dasigi et~al\mbox{.}}{2019}]%
        {dasigi:2019:acl}
\bibfield{author}{\bibinfo{person}{Pradeep Dasigi}, \bibinfo{person}{Matt
  Gardner}, \bibinfo{person}{Shikhar Murty}, \bibinfo{person}{Luke
  Zettlemoyer}, {and} \bibinfo{person}{Eduard Hovy}.}
  \bibinfo{year}{2019}\natexlab{}.
\newblock \showarticletitle{Iterative Search for Weakly Supervised Semantic
  Parsing}. In \bibinfo{booktitle}{\emph{Proceedings of the 2019 Conference of
  the North {A}merican Chapter of the Association for Computational
  Linguistics: Human Language Technologies, Volume 1 (Long and Short Papers)}}.
  \bibinfo{publisher}{Association for Computational Linguistics},
  \bibinfo{address}{Minneapolis, Minnesota}, \bibinfo{pages}{2669--2680}.
\newblock
\urldef\tempurl%
\url{https://doi.org/10.18653/v1/N19-1273}
\showDOI{\tempurl}


\bibitem[\protect\citeauthoryear{Dong and Lapata}{Dong and Lapata}{2016}]%
        {dong2016language}
\bibfield{author}{\bibinfo{person}{Li Dong} {and} \bibinfo{person}{Mirella
  Lapata}.} \bibinfo{year}{2016}\natexlab{}.
\newblock \showarticletitle{Language to Logical Form with Neural Attention}. In
  \bibinfo{booktitle}{\emph{Proceedings of the 54th Annual Meeting of the
  Association for Computational Linguistics (Volume 1: Long Papers)}}.
  \bibinfo{publisher}{Association for Computational Linguistics},
  \bibinfo{address}{Berlin, Germany}, \bibinfo{pages}{33--43}.
\newblock
\urldef\tempurl%
\url{https://doi.org/10.18653/v1/P16-1004}
\showDOI{\tempurl}


\bibitem[\protect\citeauthoryear{Dong and Lapata}{Dong and Lapata}{2018}]%
        {dong-lapata-2018-coarse}
\bibfield{author}{\bibinfo{person}{Li Dong} {and} \bibinfo{person}{Mirella
  Lapata}.} \bibinfo{year}{2018}\natexlab{}.
\newblock \showarticletitle{Coarse-to-Fine Decoding for Neural Semantic
  Parsing}. In \bibinfo{booktitle}{\emph{Proceedings of the 56th Annual Meeting
  of the Association for Computational Linguistics (Volume 1: Long Papers)}}.
  \bibinfo{publisher}{Association for Computational Linguistics},
  \bibinfo{address}{Melbourne, Australia}, \bibinfo{pages}{731--742}.
\newblock
\urldef\tempurl%
\url{https://doi.org/10.18653/v1/P18-1068}
\showDOI{\tempurl}


\bibitem[\protect\citeauthoryear{Dong, Quirk, and Lapata}{Dong
  et~al\mbox{.}}{2018}]%
        {dong2018confidence}
\bibfield{author}{\bibinfo{person}{Li Dong}, \bibinfo{person}{Chris Quirk},
  {and} \bibinfo{person}{Mirella Lapata}.} \bibinfo{year}{2018}\natexlab{}.
\newblock \showarticletitle{Confidence Modeling for Neural Semantic Parsing}.
  In \bibinfo{booktitle}{\emph{Proceedings of the 56th Annual Meeting of the
  Association for Computational Linguistics (Volume 1: Long Papers)}}.
  \bibinfo{publisher}{Association for Computational Linguistics},
  \bibinfo{address}{Melbourne, Australia}, \bibinfo{pages}{743--753}.
\newblock
\urldef\tempurl%
\url{https://doi.org/10.18653/v1/P18-1069}
\showDOI{\tempurl}


\bibitem[\protect\citeauthoryear{Feser, Chaudhuri, and Dillig}{Feser
  et~al\mbox{.}}{2015}]%
        {feser:2015:sigplan}
\bibfield{author}{\bibinfo{person}{John~K. Feser}, \bibinfo{person}{Swarat
  Chaudhuri}, {and} \bibinfo{person}{Isil Dillig}.}
  \bibinfo{year}{2015}\natexlab{}.
\newblock \showarticletitle{Synthesizing Data Structure Transformations from
  Input-Output Examples}.
\newblock \bibinfo{journal}{\emph{SIGPLAN Not.}} \bibinfo{volume}{50},
  \bibinfo{number}{6} (\bibinfo{date}{June} \bibinfo{year}{2015}),
  \bibinfo{pages}{229–239}.
\newblock
\showISSN{0362-1340}
\urldef\tempurl%
\url{https://doi.org/10.1145/2813885.2737977}
\showDOI{\tempurl}


\bibitem[\protect\citeauthoryear{Gardner and Krishnamurthy}{Gardner and
  Krishnamurthy}{2017}]%
        {gardner:2017:aaai}
\bibfield{author}{\bibinfo{person}{Matt Gardner} {and} \bibinfo{person}{J.
  Krishnamurthy}.} \bibinfo{year}{2017}\natexlab{}.
\newblock \showarticletitle{Open-Vocabulary Semantic Parsing with both
  Distributional Statistics and Formal Knowledge}. In
  \bibinfo{booktitle}{\emph{AAAI}}.
\newblock


\bibitem[\protect\citeauthoryear{Goldwasser, Reichart, Clarke, and
  Roth}{Goldwasser et~al\mbox{.}}{2011}]%
        {goldwasser-etal-2011-confidence}
\bibfield{author}{\bibinfo{person}{Dan Goldwasser}, \bibinfo{person}{Roi
  Reichart}, \bibinfo{person}{James Clarke}, {and} \bibinfo{person}{Dan Roth}.}
  \bibinfo{year}{2011}\natexlab{}.
\newblock \showarticletitle{Confidence Driven Unsupervised Semantic Parsing}.
  In \bibinfo{booktitle}{\emph{Proceedings of the 49th Annual Meeting of the
  Association for Computational Linguistics: Human Language Technologies}}.
  \bibinfo{publisher}{Association for Computational Linguistics},
  \bibinfo{address}{Portland, Oregon, USA}, \bibinfo{pages}{1486--1495}.
\newblock
\urldef\tempurl%
\url{https://www.aclweb.org/anthology/P11-1149}
\showURL{%
\tempurl}


\bibitem[\protect\citeauthoryear{Goldwasser and Roth}{Goldwasser and
  Roth}{2014}]%
        {goldwasser:2014:mljournal}
\bibfield{author}{\bibinfo{person}{Dan Goldwasser} {and} \bibinfo{person}{Dan
  Roth}.} \bibinfo{year}{2014}\natexlab{}.
\newblock \showarticletitle{Learning from natural instructions}.
\newblock \bibinfo{journal}{\emph{Machine Learning}} \bibinfo{volume}{94},
  \bibinfo{number}{2} (\bibinfo{date}{1 Feb.} \bibinfo{year}{2014}),
  \bibinfo{pages}{205--232}.
\newblock
\showISSN{0885-6125}
\urldef\tempurl%
\url{https://doi.org/10.1007/s10994-013-5407-y}
\showDOI{\tempurl}


\bibitem[\protect\citeauthoryear{Gottschlich, Solar-Lezama, Tatbul, Carbin,
  Rinard, Barzilay, Amarasinghe, Tenenbaum, and Mattson}{Gottschlich
  et~al\mbox{.}}{2018}]%
        {gottschlich2018pillars}
\bibfield{author}{\bibinfo{person}{Justin Gottschlich},
  \bibinfo{person}{Armando Solar-Lezama}, \bibinfo{person}{Nesime Tatbul},
  \bibinfo{person}{Michael Carbin}, \bibinfo{person}{Martin Rinard},
  \bibinfo{person}{Regina Barzilay}, \bibinfo{person}{Saman Amarasinghe},
  \bibinfo{person}{Joshua~B Tenenbaum}, {and} \bibinfo{person}{Tim Mattson}.}
  \bibinfo{year}{2018}\natexlab{}.
\newblock \bibinfo{title}{The Three Pillars of Machine Programming}.
\newblock
\newblock
\showeprint[arxiv]{1803.07244}~[cs.AI]


\bibitem[\protect\citeauthoryear{Gu, Lu, Li, and Li}{Gu et~al\mbox{.}}{2016}]%
        {gu:2016:acl}
\bibfield{author}{\bibinfo{person}{Jiatao Gu}, \bibinfo{person}{Zhengdong Lu},
  \bibinfo{person}{Hang Li}, {and} \bibinfo{person}{Victor~O.K. Li}.}
  \bibinfo{year}{2016}\natexlab{}.
\newblock \showarticletitle{Incorporating Copying Mechanism in
  Sequence-to-Sequence Learning}. In \bibinfo{booktitle}{\emph{Proceedings of
  the 54th Annual Meeting of the Association for Computational Linguistics
  (Volume 1: Long Papers)}}. \bibinfo{publisher}{Association for Computational
  Linguistics}, \bibinfo{address}{Berlin, Germany},
  \bibinfo{pages}{1631--1640}.
\newblock
\urldef\tempurl%
\url{https://doi.org/10.18653/v1/P16-1154}
\showDOI{\tempurl}


\bibitem[\protect\citeauthoryear{Gulwani}{Gulwani}{2011}]%
        {gulwani:2011:sigplan}
\bibfield{author}{\bibinfo{person}{Sumit Gulwani}.}
  \bibinfo{year}{2011}\natexlab{}.
\newblock \showarticletitle{Automating String Processing in Spreadsheets Using
  Input-Output Examples}.
\newblock \bibinfo{journal}{\emph{SIGPLAN Not.}} \bibinfo{volume}{46},
  \bibinfo{number}{1} (\bibinfo{date}{Jan.} \bibinfo{year}{2011}),
  \bibinfo{pages}{317–330}.
\newblock
\showISSN{0362-1340}
\urldef\tempurl%
\url{https://doi.org/10.1145/1925844.1926423}
\showDOI{\tempurl}


\bibitem[\protect\citeauthoryear{Gulwani and Jain}{Gulwani and Jain}{2017}]%
        {gulwani:2017:aplas}
\bibfield{author}{\bibinfo{person}{Sumit Gulwani} {and}
  \bibinfo{person}{Prateek Jain}.} \bibinfo{year}{2017}\natexlab{}.
\newblock \showarticletitle{Programming by Examples: {PL} Meets {ML}}. In
  \bibinfo{booktitle}{\emph{Programming Languages and Systems - 15th Asian
  Symposium, {APLAS} 2017, Suzhou, China, November 27-29, 2017, Proceedings}}
  \emph{(\bibinfo{series}{Lecture Notes in Computer Science},
  Vol.~\bibinfo{volume}{10695})},
  \bibfield{editor}{\bibinfo{person}{Bor{-}Yuh~Evan Chang}} (Ed.).
  \bibinfo{publisher}{Springer}, \bibinfo{pages}{3--20}.
\newblock
\urldef\tempurl%
\url{https://doi.org/10.1007/978-3-319-71237-6\_1}
\showDOI{\tempurl}


\bibitem[\protect\citeauthoryear{Gupta, Lin, Roth, Singh, and Gardner}{Gupta
  et~al\mbox{.}}{2020}]%
        {gupta:2020:iclr}
\bibfield{author}{\bibinfo{person}{Nitish Gupta}, \bibinfo{person}{Kevin Lin},
  \bibinfo{person}{Dan Roth}, \bibinfo{person}{Sameer Singh}, {and}
  \bibinfo{person}{Matt Gardner}.} \bibinfo{year}{2020}\natexlab{}.
\newblock \showarticletitle{Neural Module Networks for Reasoning over Text}. In
  \bibinfo{booktitle}{\emph{International Conference on Learning
  Representations}}.
\newblock


\bibitem[\protect\citeauthoryear{Hendrix, Sacerdoti, Sagalowicz, and
  Slocum}{Hendrix et~al\mbox{.}}{1978}]%
        {hendrix:1978:acm}
\bibfield{author}{\bibinfo{person}{Gary~G. Hendrix}, \bibinfo{person}{Earl~D.
  Sacerdoti}, \bibinfo{person}{Daniel Sagalowicz}, {and}
  \bibinfo{person}{Jonathan Slocum}.} \bibinfo{year}{1978}\natexlab{}.
\newblock \showarticletitle{Developing a Natural Language Interface to Complex
  Data}.
\newblock \bibinfo{journal}{\emph{ACM Trans. Database Syst.}}
  \bibinfo{volume}{3}, \bibinfo{number}{2} (\bibinfo{date}{June}
  \bibinfo{year}{1978}), \bibinfo{pages}{105–147}.
\newblock
\showISSN{0362-5915}
\urldef\tempurl%
\url{https://doi.org/10.1145/320251.320253}
\showDOI{\tempurl}


\bibitem[\protect\citeauthoryear{Hockenmaier and Steedman}{Hockenmaier and
  Steedman}{2007}]%
        {hockenmaier:2007:computationallinguistics}
\bibfield{author}{\bibinfo{person}{Julia Hockenmaier} {and}
  \bibinfo{person}{Mark Steedman}.} \bibinfo{year}{2007}\natexlab{}.
\newblock \showarticletitle{{CCG}bank: A Corpus of {CCG} Derivations and
  Dependency Structures Extracted from the {P}enn {T}reebank}.
\newblock \bibinfo{journal}{\emph{Computational Linguistics}}
  \bibinfo{volume}{33}, \bibinfo{number}{3} (\bibinfo{year}{2007}),
  \bibinfo{pages}{355--396}.
\newblock
\urldef\tempurl%
\url{https://doi.org/10.1162/coli.2007.33.3.355}
\showDOI{\tempurl}


\bibitem[\protect\citeauthoryear{Iyer, Sun, Wang, and Gottschlich}{Iyer
  et~al\mbox{.}}{2020}]%
        {iyer2020software}
\bibfield{author}{\bibinfo{person}{Roshni~G. Iyer}, \bibinfo{person}{Yizhou
  Sun}, \bibinfo{person}{Wei Wang}, {and} \bibinfo{person}{Justin
  Gottschlich}.} \bibinfo{year}{2020}\natexlab{}.
\newblock \bibinfo{title}{Software Language Comprehension using a
  Program-Derived Semantics Graph}.
\newblock
\newblock
\showeprint[arxiv]{2004.00768}~[cs.AI]


\bibitem[\protect\citeauthoryear{Iyer, Cheung, and Zettlemoyer}{Iyer
  et~al\mbox{.}}{2019}]%
        {iyer2019learning}
\bibfield{author}{\bibinfo{person}{Srinivasan Iyer}, \bibinfo{person}{Alvin
  Cheung}, {and} \bibinfo{person}{Luke Zettlemoyer}.}
  \bibinfo{year}{2019}\natexlab{}.
\newblock \showarticletitle{Learning Programmatic Idioms for Scalable Semantic
  Parsing}. In \bibinfo{booktitle}{\emph{Proceedings of the 2019 Conference on
  Empirical Methods in Natural Language Processing and the 9th International
  Joint Conference on Natural Language Processing, {EMNLP-IJCNLP} 2019, Hong
  Kong, China, November 3-7, 2019}}, \bibfield{editor}{\bibinfo{person}{Kentaro
  Inui}, \bibinfo{person}{Jing Jiang}, \bibinfo{person}{Vincent Ng}, {and}
  \bibinfo{person}{Xiaojun Wan}} (Eds.). \bibinfo{publisher}{Association for
  Computational Linguistics}, \bibinfo{pages}{5425--5434}.
\newblock
\urldef\tempurl%
\url{https://doi.org/10.18653/v1/D19-1545}
\showDOI{\tempurl}


\bibitem[\protect\citeauthoryear{Iyer, Konstas, Cheung, Krishnamurthy, and
  Zettlemoyer}{Iyer et~al\mbox{.}}{2017}]%
        {iyer:2017:acl}
\bibfield{author}{\bibinfo{person}{Srinivasan Iyer}, \bibinfo{person}{Ioannis
  Konstas}, \bibinfo{person}{Alvin Cheung}, \bibinfo{person}{Jayant
  Krishnamurthy}, {and} \bibinfo{person}{Luke Zettlemoyer}.}
  \bibinfo{year}{2017}\natexlab{}.
\newblock \showarticletitle{Learning a Neural Semantic Parser from User
  Feedback}. In \bibinfo{booktitle}{\emph{Proceedings of the 55th Annual
  Meeting of the Association for Computational Linguistics (Volume 1: Long
  Papers)}}. \bibinfo{publisher}{Association for Computational Linguistics},
  \bibinfo{address}{Vancouver, Canada}, \bibinfo{pages}{963--973}.
\newblock
\urldef\tempurl%
\url{https://doi.org/10.18653/v1/P17-1089}
\showDOI{\tempurl}


\bibitem[\protect\citeauthoryear{Iyer, Konstas, Cheung, and Zettlemoyer}{Iyer
  et~al\mbox{.}}{2018}]%
        {iyer-etal-2018-mapping}
\bibfield{author}{\bibinfo{person}{Srinivasan Iyer}, \bibinfo{person}{Ioannis
  Konstas}, \bibinfo{person}{Alvin Cheung}, {and} \bibinfo{person}{Luke
  Zettlemoyer}.} \bibinfo{year}{2018}\natexlab{}.
\newblock \showarticletitle{Mapping Language to Code in Programmatic Context}.
  In \bibinfo{booktitle}{\emph{Proceedings of the 2018 Conference on Empirical
  Methods in Natural Language Processing}}. \bibinfo{publisher}{Association for
  Computational Linguistics}, \bibinfo{address}{Brussels, Belgium},
  \bibinfo{pages}{1643--1652}.
\newblock
\urldef\tempurl%
\url{https://doi.org/10.18653/v1/D18-1192}
\showDOI{\tempurl}


\bibitem[\protect\citeauthoryear{Johnson}{Johnson}{1984}]%
        {johnson_1984}
\bibfield{author}{\bibinfo{person}{Tim Johnson}.}
  \bibinfo{year}{1984}\natexlab{}.
\newblock \showarticletitle{Natural Language Computing: The Commercial
  Applications}.
\newblock \bibinfo{journal}{\emph{The Knowledge Engineering Review}}
  \bibinfo{volume}{1}, \bibinfo{number}{3} (\bibinfo{year}{1984}),
  \bibinfo{pages}{11–23}.
\newblock
\urldef\tempurl%
\url{https://doi.org/10.1017/S0269888900000588}
\showDOI{\tempurl}


\bibitem[\protect\citeauthoryear{Kamil, Cheung, Itzhaky, and
  Solar-Lezama}{Kamil et~al\mbox{.}}{2016}]%
        {kamil:2016:pldi}
\bibfield{author}{\bibinfo{person}{Shoaib Kamil}, \bibinfo{person}{Alvin
  Cheung}, \bibinfo{person}{Shachar Itzhaky}, {and} \bibinfo{person}{Armando
  Solar-Lezama}.} \bibinfo{year}{2016}\natexlab{}.
\newblock \showarticletitle{Verified Lifting of Stencil Computations}.
\newblock \bibinfo{journal}{\emph{SIGPLAN Not.}} \bibinfo{volume}{51},
  \bibinfo{number}{6} (\bibinfo{date}{June} \bibinfo{year}{2016}),
  \bibinfo{pages}{711–726}.
\newblock
\showISSN{0362-1340}
\urldef\tempurl%
\url{https://doi.org/10.1145/2980983.2908117}
\showDOI{\tempurl}


\bibitem[\protect\citeauthoryear{Kate, Wong, and Mooney}{Kate
  et~al\mbox{.}}{2005}]%
        {kate:2005:ncai}
\bibfield{author}{\bibinfo{person}{Rohit~J. Kate}, \bibinfo{person}{Yuk~Wah
  Wong}, {and} \bibinfo{person}{Raymond~J. Mooney}.}
  \bibinfo{year}{2005}\natexlab{}.
\newblock \showarticletitle{Learning to Transform Natural to Formal Languages}.
  In \bibinfo{booktitle}{\emph{Proceedings of the 20th National Conference on
  Artificial Intelligence - Volume 3}} (Pittsburgh, Pennsylvania)
  \emph{(\bibinfo{series}{AAAI'05})}. \bibinfo{publisher}{AAAI Press},
  \bibinfo{pages}{1062–1068}.
\newblock
\showISBNx{157735236x}


\bibitem[\protect\citeauthoryear{Krishnamurthy, Dasigi, and
  Gardner}{Krishnamurthy et~al\mbox{.}}{2017}]%
        {krishnamurthy:2017:empiricalmethodsnl}
\bibfield{author}{\bibinfo{person}{Jayant Krishnamurthy},
  \bibinfo{person}{Pradeep Dasigi}, {and} \bibinfo{person}{Matt Gardner}.}
  \bibinfo{year}{2017}\natexlab{}.
\newblock \showarticletitle{Neural Semantic Parsing with Type Constraints for
  Semi-Structured Tables}. In \bibinfo{booktitle}{\emph{Proceedings of the 2017
  Conference on Empirical Methods in Natural Language Processing}}.
  \bibinfo{publisher}{Association for Computational Linguistics},
  \bibinfo{address}{Copenhagen, Denmark}, \bibinfo{pages}{1516--1526}.
\newblock
\urldef\tempurl%
\url{https://doi.org/10.18653/v1/D17-1160}
\showDOI{\tempurl}


\bibitem[\protect\citeauthoryear{Krishnamurthy and Mitchell}{Krishnamurthy and
  Mitchell}{2015}]%
        {krishnamurthy:2015:tacl}
\bibfield{author}{\bibinfo{person}{Jayant Krishnamurthy} {and}
  \bibinfo{person}{Tom~M. Mitchell}.} \bibinfo{year}{2015}\natexlab{}.
\newblock \showarticletitle{Learning a Compositional Semantics for {F}reebase
  with an Open Predicate Vocabulary}.
\newblock \bibinfo{journal}{\emph{Transactions of the Association for
  Computational Linguistics}}  \bibinfo{volume}{3} (\bibinfo{year}{2015}),
  \bibinfo{pages}{257--270}.
\newblock
\urldef\tempurl%
\url{https://doi.org/10.1162/tacl_a_00137}
\showDOI{\tempurl}


\bibitem[\protect\citeauthoryear{Kulal, Pasupat, Chandra, Lee, Padon, Aiken,
  and Liang}{Kulal et~al\mbox{.}}{2019}]%
        {kulal:2019:neurips}
\bibfield{author}{\bibinfo{person}{S. Kulal}, \bibinfo{person}{Panupong
  Pasupat}, \bibinfo{person}{K. Chandra}, \bibinfo{person}{Mina Lee},
  \bibinfo{person}{Oded Padon}, \bibinfo{person}{A. Aiken}, {and}
  \bibinfo{person}{Percy Liang}.} \bibinfo{year}{2019}\natexlab{}.
\newblock \showarticletitle{SPoC: Search-based Pseudocode to Code}. In
  \bibinfo{booktitle}{\emph{NeurIPS}}.
\newblock


\bibitem[\protect\citeauthoryear{Kwiatkowski, Choi, Artzi, and
  Zettlemoyer}{Kwiatkowski et~al\mbox{.}}{2013}]%
        {kwiatkowski-etal-2013-scaling}
\bibfield{author}{\bibinfo{person}{Tom Kwiatkowski}, \bibinfo{person}{Eunsol
  Choi}, \bibinfo{person}{Yoav Artzi}, {and} \bibinfo{person}{Luke
  Zettlemoyer}.} \bibinfo{year}{2013}\natexlab{}.
\newblock \showarticletitle{Scaling Semantic Parsers with On-the-Fly Ontology
  Matching}. In \bibinfo{booktitle}{\emph{Proceedings of the 2013 Conference on
  Empirical Methods in Natural Language Processing}}.
  \bibinfo{publisher}{Association for Computational Linguistics},
  \bibinfo{address}{Seattle, Washington, USA}, \bibinfo{pages}{1545--1556}.
\newblock
\urldef\tempurl%
\url{https://www.aclweb.org/anthology/D13-1161}
\showURL{%
\tempurl}


\bibitem[\protect\citeauthoryear{Lewis and Steedman}{Lewis and
  Steedman}{2013}]%
        {lewis:2013:tacl}
\bibfield{author}{\bibinfo{person}{Mike Lewis} {and} \bibinfo{person}{Mark
  Steedman}.} \bibinfo{year}{2013}\natexlab{}.
\newblock \showarticletitle{Combined Distributional and Logical Semantics}.
\newblock \bibinfo{journal}{\emph{Transactions of the Association for
  Computational Linguistics}}  \bibinfo{volume}{1} (\bibinfo{year}{2013}),
  \bibinfo{pages}{179--192}.
\newblock
\urldef\tempurl%
\url{https://doi.org/10.1162/tacl_a_00219}
\showDOI{\tempurl}


\bibitem[\protect\citeauthoryear{Liang, Jordan, and Klein}{Liang
  et~al\mbox{.}}{2011}]%
        {liang:2011:acl}
\bibfield{author}{\bibinfo{person}{Percy Liang}, \bibinfo{person}{Michael
  Jordan}, {and} \bibinfo{person}{Dan Klein}.} \bibinfo{year}{2011}\natexlab{}.
\newblock \showarticletitle{{Learning Dependency-Based Compositional
  Semantics}}. In \bibinfo{booktitle}{\emph{Proceedings of the 49th Annual
  Meeting of the Association for Computational Linguistics: Human Language
  Technologies}}. \bibinfo{publisher}{Association for Computational
  Linguistics}, \bibinfo{address}{Portland, Oregon, USA},
  \bibinfo{pages}{590--599}.
\newblock
\urldef\tempurl%
\url{https://www.aclweb.org/anthology/P11-1060}
\showURL{%
\tempurl}


\bibitem[\protect\citeauthoryear{Lieberman}{Lieberman}{2001}]%
        {lieberman:2001:book}
\bibfield{author}{\bibinfo{person}{H. Lieberman}.}
  \bibinfo{year}{2001}\natexlab{}.
\newblock \showarticletitle{Your wish is my command: Programming by example}.
\newblock  (\bibinfo{date}{01} \bibinfo{year}{2001}).
\newblock


\bibitem[\protect\citeauthoryear{Ling, Blunsom, Grefenstette, Hermann,
  Ko{\v{c}}isk{\'y}, Wang, and Senior}{Ling et~al\mbox{.}}{2016}]%
        {ling2016latent}
\bibfield{author}{\bibinfo{person}{Wang Ling}, \bibinfo{person}{Phil Blunsom},
  \bibinfo{person}{Edward Grefenstette}, \bibinfo{person}{Karl~Moritz Hermann},
  \bibinfo{person}{Tom{\'a}{\v{s}} Ko{\v{c}}isk{\'y}}, \bibinfo{person}{Fumin
  Wang}, {and} \bibinfo{person}{Andrew Senior}.}
  \bibinfo{year}{2016}\natexlab{}.
\newblock \showarticletitle{Latent Predictor Networks for Code Generation}. In
  \bibinfo{booktitle}{\emph{Proceedings of the 54th Annual Meeting of the
  Association for Computational Linguistics (Volume 1: Long Papers)}}.
  \bibinfo{publisher}{Association for Computational Linguistics},
  \bibinfo{address}{Berlin, Germany}, \bibinfo{pages}{599--609}.
\newblock
\urldef\tempurl%
\url{https://doi.org/10.18653/v1/P16-1057}
\showDOI{\tempurl}


\bibitem[\protect\citeauthoryear{Lockemann and Thompson}{Lockemann and
  Thompson}{1969}]%
        {lockemann-thompson-1969-rapidly}
\bibfield{author}{\bibinfo{person}{Peter~C. Lockemann} {and}
  \bibinfo{person}{Frederick~B. Thompson}.} \bibinfo{year}{1969}\natexlab{}.
\newblock \showarticletitle{A Rapidly Extensible Language System (The {REL}
  Language Processor)}. In \bibinfo{booktitle}{\emph{{I}nternational
  {C}onference on {C}omputational {L}inguistics {COLING} 1969: Preprint No.
  34}}. \bibinfo{address}{S{\aa}nga S{\"a}by, Sweden}.
\newblock
\urldef\tempurl%
\url{https://www.aclweb.org/anthology/C69-3401}
\showURL{%
\tempurl}


\bibitem[\protect\citeauthoryear{Luan, Yang, Barnaby, Sen, and Chandra}{Luan
  et~al\mbox{.}}{2019}]%
        {luan:2019:acm}
\bibfield{author}{\bibinfo{person}{Sifei Luan}, \bibinfo{person}{Di Yang},
  \bibinfo{person}{Celeste Barnaby}, \bibinfo{person}{Koushik Sen}, {and}
  \bibinfo{person}{Satish Chandra}.} \bibinfo{year}{2019}\natexlab{}.
\newblock \showarticletitle{Aroma: Code Recommendation via Structural Code
  Search}.
\newblock \bibinfo{journal}{\emph{Proc. ACM Program. Lang.}}
  \bibinfo{volume}{3}, \bibinfo{number}{OOPSLA}, Article
  \bibinfo{articleno}{152} (\bibinfo{date}{Oct.} \bibinfo{year}{2019}),
  \bibinfo{numpages}{28}~pages.
\newblock
\urldef\tempurl%
\url{https://doi.org/10.1145/3360578}
\showDOI{\tempurl}


\bibitem[\protect\citeauthoryear{Machines, Andreas, Bufe, Burkett, Chen,
  Clausman, Crawford, Crim, DeLoach, Dorner, Eisner, Fang, Guo, Hall, Hayes,
  Hill, Ho, Iwaszuk, Jha, Klein, Krishnamurthy, Lanman, Liang, Lin, Lintsbakh,
  McGovern, Nisnevich, Pauls, Petters, Read, Roth, Roy, Rusak, Short, Slomin,
  Snyder, Striplin, Su, Tellman, Thomson, Vorobev, Witoszko, Wolfe, Wray,
  Zhang, and Zotov}{Machines et~al\mbox{.}}{2020}]%
        {machines2020task-oriented}
\bibfield{author}{\bibinfo{person}{Semantic Machines}, \bibinfo{person}{Jacob
  Andreas}, \bibinfo{person}{John Bufe}, \bibinfo{person}{David Burkett},
  \bibinfo{person}{Charles Chen}, \bibinfo{person}{Josh Clausman},
  \bibinfo{person}{Jean Crawford}, \bibinfo{person}{Kate Crim},
  \bibinfo{person}{Jordan DeLoach}, \bibinfo{person}{Leah Dorner},
  \bibinfo{person}{Jason Eisner}, \bibinfo{person}{Hao Fang},
  \bibinfo{person}{Alan Guo}, \bibinfo{person}{David Hall},
  \bibinfo{person}{Kristin Hayes}, \bibinfo{person}{Kellie Hill},
  \bibinfo{person}{Diana Ho}, \bibinfo{person}{Wendy Iwaszuk},
  \bibinfo{person}{Smriti Jha}, \bibinfo{person}{Dan Klein},
  \bibinfo{person}{Jayant Krishnamurthy}, \bibinfo{person}{Theo Lanman},
  \bibinfo{person}{Percy Liang}, \bibinfo{person}{Christopher~H Lin},
  \bibinfo{person}{Ilya Lintsbakh}, \bibinfo{person}{Andy McGovern},
  \bibinfo{person}{Aleksandr Nisnevich}, \bibinfo{person}{Adam Pauls},
  \bibinfo{person}{Dmitrij Petters}, \bibinfo{person}{Brent Read},
  \bibinfo{person}{Dan Roth}, \bibinfo{person}{Subhro Roy},
  \bibinfo{person}{Jesse Rusak}, \bibinfo{person}{Beth Short},
  \bibinfo{person}{Div Slomin}, \bibinfo{person}{Ben Snyder},
  \bibinfo{person}{Stephon Striplin}, \bibinfo{person}{Yu Su},
  \bibinfo{person}{Zachary Tellman}, \bibinfo{person}{Sam Thomson},
  \bibinfo{person}{Andrei Vorobev}, \bibinfo{person}{Izabela Witoszko},
  \bibinfo{person}{Jason Wolfe}, \bibinfo{person}{Abby Wray},
  \bibinfo{person}{Yuchen Zhang}, {and} \bibinfo{person}{Alexander Zotov}.}
  \bibinfo{year}{2020}\natexlab{}.
\newblock \showarticletitle{Task-Oriented Dialogue as Dataflow Synthesis}.
\newblock \bibinfo{journal}{\emph{Transactions of the Association for
  Computational Linguistics}}  \bibinfo{volume}{8} (\bibinfo{date}{September}
  \bibinfo{year}{2020}).
\newblock
\urldef\tempurl%
\url{https://www.microsoft.com/en-us/research/publication/task-oriented-dialogue-as-dataflow-synthesis/}
\showURL{%
\tempurl}


\bibitem[\protect\citeauthoryear{Mandal, Anderson, Turek, Gottschlich, Zhou,
  and Muzahid}{Mandal et~al\mbox{.}}{2020}]%
        {mandal:2021:mlsys}
\bibfield{author}{\bibinfo{person}{Shantanu Mandal}, \bibinfo{person}{T.
  Anderson}, \bibinfo{person}{Javier Turek}, \bibinfo{person}{Justin~Emile
  Gottschlich}, \bibinfo{person}{Sheng-Tian Zhou}, {and} \bibinfo{person}{A.
  Muzahid}.} \bibinfo{year}{2020}\natexlab{}.
\newblock \showarticletitle{Learning Fitness Functions for Machine
  Programming.}
\newblock \bibinfo{journal}{\emph{arXiv: Neural and Evolutionary Computing}}
  (\bibinfo{year}{2020}).
\newblock


\bibitem[\protect\citeauthoryear{Manna and Waldinger}{Manna and
  Waldinger}{1980}]%
        {zohar:1980:acm}
\bibfield{author}{\bibinfo{person}{Zohar Manna} {and} \bibinfo{person}{Richard
  Waldinger}.} \bibinfo{year}{1980}\natexlab{}.
\newblock \showarticletitle{A Deductive Approach to Program Synthesis}.
\newblock \bibinfo{journal}{\emph{ACM Trans. Program. Lang. Syst.}}
  \bibinfo{volume}{2}, \bibinfo{number}{1} (\bibinfo{date}{Jan.}
  \bibinfo{year}{1980}), \bibinfo{pages}{90–121}.
\newblock
\showISSN{0164-0925}
\urldef\tempurl%
\url{https://doi.org/10.1145/357084.357090}
\showDOI{\tempurl}


\bibitem[\protect\citeauthoryear{Marcus, Negi, Mao, Tatbul, Alizadeh, and
  Kraska}{Marcus et~al\mbox{.}}{2020}]%
        {marcus2020bao}
\bibfield{author}{\bibinfo{person}{Ryan Marcus}, \bibinfo{person}{Parimarjan
  Negi}, \bibinfo{person}{Hongzi Mao}, \bibinfo{person}{Nesime Tatbul},
  \bibinfo{person}{Mohammad Alizadeh}, {and} \bibinfo{person}{Tim Kraska}.}
  \bibinfo{year}{2020}\natexlab{}.
\newblock \bibinfo{title}{Bao: Learning to Steer Query Optimizers}.
\newblock
\newblock
\showeprint[arxiv]{2004.03814}~[cs.DB]


\bibitem[\protect\citeauthoryear{Marcus, Negi, Mao, Zhang, Alizadeh, Kraska,
  Papaemmanouil, and Tatbul}{Marcus et~al\mbox{.}}{2019}]%
        {marcus:2019:vldb}
\bibfield{author}{\bibinfo{person}{Ryan Marcus}, \bibinfo{person}{Parimarjan
  Negi}, \bibinfo{person}{Hongzi Mao}, \bibinfo{person}{Chi Zhang},
  \bibinfo{person}{Mohammad Alizadeh}, \bibinfo{person}{Tim Kraska},
  \bibinfo{person}{Olga Papaemmanouil}, {and} \bibinfo{person}{Nesime Tatbul}.}
  \bibinfo{year}{2019}\natexlab{}.
\newblock \showarticletitle{Neo: A Learned Query Optimizer}.
\newblock \bibinfo{journal}{\emph{Proc. VLDB Endow.}} \bibinfo{volume}{12},
  \bibinfo{number}{11} (\bibinfo{date}{July} \bibinfo{year}{2019}),
  \bibinfo{pages}{1705–1718}.
\newblock
\showISSN{2150-8097}
\urldef\tempurl%
\url{https://doi.org/10.14778/3342263.3342644}
\showDOI{\tempurl}


\bibitem[\protect\citeauthoryear{Menon, Tamuz, Gulwani, Lampson, and
  Kalai}{Menon et~al\mbox{.}}{2013}]%
        {pmlr-v28-menon13}
\bibfield{author}{\bibinfo{person}{Aditya Menon}, \bibinfo{person}{Omer Tamuz},
  \bibinfo{person}{Sumit Gulwani}, \bibinfo{person}{Butler Lampson}, {and}
  \bibinfo{person}{Adam Kalai}.} \bibinfo{year}{2013}\natexlab{}.
\newblock \showarticletitle{A Machine Learning Framework for Programming by
  Example}. In \bibinfo{booktitle}{\emph{Proceedings of the 30th International
  Conference on Machine Learning}} \emph{(\bibinfo{series}{Proceedings of
  Machine Learning Research}, Vol.~\bibinfo{volume}{28})},
  \bibfield{editor}{\bibinfo{person}{Sanjoy Dasgupta} {and}
  \bibinfo{person}{David McAllester}} (Eds.). \bibinfo{publisher}{PMLR},
  \bibinfo{address}{Atlanta, Georgia, USA}, \bibinfo{pages}{187--195}.
\newblock
\urldef\tempurl%
\url{http://proceedings.mlr.press/v28/menon13.html}
\showURL{%
\tempurl}


\bibitem[\protect\citeauthoryear{Nye, Hewitt, Tenenbaum, and Solar-Lezama}{Nye
  et~al\mbox{.}}{2019}]%
        {nye2019learning}
\bibfield{author}{\bibinfo{person}{Maxwell Nye}, \bibinfo{person}{Luke Hewitt},
  \bibinfo{person}{Joshua Tenenbaum}, {and} \bibinfo{person}{Armando
  Solar-Lezama}.} \bibinfo{year}{2019}\natexlab{}.
\newblock \bibinfo{title}{Learning to Infer Program Sketches}.
\newblock , \bibinfo{numpages}{4861--4870}~pages.
\newblock
\urldef\tempurl%
\url{http://proceedings.mlr.press/v97/nye19a.html}
\showURL{%
\tempurl}


\bibitem[\protect\citeauthoryear{Oepen, Kuhlmann, Miyao, Zeman, Flickinger,
  Haji{\v{c}}, Ivanova, and Zhang}{Oepen et~al\mbox{.}}{2014}]%
        {oepen-etal-2014-semeval}
\bibfield{author}{\bibinfo{person}{Stephan Oepen}, \bibinfo{person}{Marco
  Kuhlmann}, \bibinfo{person}{Yusuke Miyao}, \bibinfo{person}{Daniel Zeman},
  \bibinfo{person}{Dan Flickinger}, \bibinfo{person}{Jan Haji{\v{c}}},
  \bibinfo{person}{Angelina Ivanova}, {and} \bibinfo{person}{Yi Zhang}.}
  \bibinfo{year}{2014}\natexlab{}.
\newblock \showarticletitle{{S}em{E}val 2014 Task 8: Broad-Coverage Semantic
  Dependency Parsing}. In \bibinfo{booktitle}{\emph{Proceedings of the 8th
  International Workshop on Semantic Evaluation ({S}em{E}val 2014)}}.
  \bibinfo{publisher}{Association for Computational Linguistics},
  \bibinfo{address}{Dublin, Ireland}, \bibinfo{pages}{63--72}.
\newblock
\urldef\tempurl%
\url{https://doi.org/10.3115/v1/S14-2008}
\showDOI{\tempurl}


\bibitem[\protect\citeauthoryear{Oren, Herzig, Gupta, Gardner, and Berant}{Oren
  et~al\mbox{.}}{2020}]%
        {oren:2020:emnlp}
\bibfield{author}{\bibinfo{person}{Inbar Oren}, \bibinfo{person}{Jonathan
  Herzig}, \bibinfo{person}{Nitish Gupta}, \bibinfo{person}{Matt Gardner},
  {and} \bibinfo{person}{Jonathan Berant}.} \bibinfo{year}{2020}\natexlab{}.
\newblock \showarticletitle{Improving Compositional Generalization in Semantic
  Parsing}. In \bibinfo{booktitle}{\emph{Findings of the Association for
  Computational Linguistics: EMNLP 2020}}. \bibinfo{publisher}{Association for
  Computational Linguistics}, \bibinfo{address}{Online},
  \bibinfo{pages}{2482--2495}.
\newblock
\urldef\tempurl%
\url{https://doi.org/10.18653/v1/2020.findings-emnlp.225}
\showDOI{\tempurl}


\bibitem[\protect\citeauthoryear{Papineni, Roukos, Ward, and Zhu}{Papineni
  et~al\mbox{.}}{2002}]%
        {papineni-etal-2002-bleu}
\bibfield{author}{\bibinfo{person}{Kishore Papineni}, \bibinfo{person}{Salim
  Roukos}, \bibinfo{person}{Todd Ward}, {and} \bibinfo{person}{Wei-Jing Zhu}.}
  \bibinfo{year}{2002}\natexlab{}.
\newblock \showarticletitle{{B}leu: a Method for Automatic Evaluation of
  Machine Translation}. In \bibinfo{booktitle}{\emph{Proceedings of the 40th
  Annual Meeting of the Association for Computational Linguistics}}.
  \bibinfo{publisher}{Association for Computational Linguistics},
  \bibinfo{address}{Philadelphia, Pennsylvania, USA},
  \bibinfo{pages}{311--318}.
\newblock
\urldef\tempurl%
\url{https://doi.org/10.3115/1073083.1073135}
\showDOI{\tempurl}


\bibitem[\protect\citeauthoryear{Pasupat and Liang}{Pasupat and Liang}{2016}]%
        {pasupat-liang-2016-inferring}
\bibfield{author}{\bibinfo{person}{Panupong Pasupat} {and}
  \bibinfo{person}{Percy Liang}.} \bibinfo{year}{2016}\natexlab{}.
\newblock \showarticletitle{Inferring Logical Forms From Denotations}. In
  \bibinfo{booktitle}{\emph{Proceedings of the 54th Annual Meeting of the
  Association for Computational Linguistics (Volume 1: Long Papers)}}.
  \bibinfo{publisher}{Association for Computational Linguistics},
  \bibinfo{address}{Berlin, Germany}, \bibinfo{pages}{23--32}.
\newblock
\urldef\tempurl%
\url{https://doi.org/10.18653/v1/P16-1003}
\showDOI{\tempurl}


\bibitem[\protect\citeauthoryear{Perelman, Gulwani, Grossman, and
  Provost}{Perelman et~al\mbox{.}}{2014}]%
        {perelman:2014:sigplan}
\bibfield{author}{\bibinfo{person}{Daniel Perelman}, \bibinfo{person}{Sumit
  Gulwani}, \bibinfo{person}{Dan Grossman}, {and} \bibinfo{person}{Peter
  Provost}.} \bibinfo{year}{2014}\natexlab{}.
\newblock \showarticletitle{Test-Driven Synthesis}.
\newblock \bibinfo{journal}{\emph{SIGPLAN Not.}} \bibinfo{volume}{49},
  \bibinfo{number}{6} (\bibinfo{date}{June} \bibinfo{year}{2014}),
  \bibinfo{pages}{408–418}.
\newblock
\showISSN{0362-1340}
\urldef\tempurl%
\url{https://doi.org/10.1145/2666356.2594297}
\showDOI{\tempurl}


\bibitem[\protect\citeauthoryear{Poon}{Poon}{2013}]%
        {poon-2013-grounded}
\bibfield{author}{\bibinfo{person}{Hoifung Poon}.}
  \bibinfo{year}{2013}\natexlab{}.
\newblock \showarticletitle{Grounded Unsupervised Semantic Parsing}. In
  \bibinfo{booktitle}{\emph{Proceedings of the 51st Annual Meeting of the
  Association for Computational Linguistics (Volume 1: Long Papers)}}.
  \bibinfo{publisher}{Association for Computational Linguistics},
  \bibinfo{address}{Sofia, Bulgaria}, \bibinfo{pages}{933--943}.
\newblock
\urldef\tempurl%
\url{https://www.aclweb.org/anthology/P13-1092}
\showURL{%
\tempurl}


\bibitem[\protect\citeauthoryear{Poon and Domingos}{Poon and Domingos}{2009}]%
        {poon:2009:emnlp}
\bibfield{author}{\bibinfo{person}{Hoifung Poon} {and} \bibinfo{person}{Pedro
  Domingos}.} \bibinfo{year}{2009}\natexlab{}.
\newblock \showarticletitle{Unsupervised Semantic Parsing}. In
  \bibinfo{booktitle}{\emph{Proceedings of the 2009 Conference on Empirical
  Methods in Natural Language Processing: Volume 1 - Volume 1}} (Singapore)
  \emph{(\bibinfo{series}{EMNLP '09})}. \bibinfo{publisher}{Association for
  Computational Linguistics}, \bibinfo{address}{USA}, \bibinfo{pages}{1–10}.
\newblock
\showISBNx{9781932432596}


\bibitem[\protect\citeauthoryear{Pu, Miranda, Solar-Lezama, and Kaelbling}{Pu
  et~al\mbox{.}}{2018}]%
        {pu:2018:pmlr}
\bibfield{author}{\bibinfo{person}{Yewen Pu}, \bibinfo{person}{Zachery
  Miranda}, \bibinfo{person}{Armando Solar-Lezama}, {and}
  \bibinfo{person}{Leslie Kaelbling}.} \bibinfo{year}{2018}\natexlab{}.
\newblock \showarticletitle{Selecting Representative Examples for Program
  Synthesis}. In \bibinfo{booktitle}{\emph{Proceedings of the 35th
  International Conference on Machine Learning}}
  \emph{(\bibinfo{series}{Proceedings of Machine Learning Research},
  Vol.~\bibinfo{volume}{80})}, \bibfield{editor}{\bibinfo{person}{Jennifer Dy}
  {and} \bibinfo{person}{Andreas Krause}} (Eds.). \bibinfo{publisher}{PMLR},
  \bibinfo{pages}{4161--4170}.
\newblock
\urldef\tempurl%
\url{http://proceedings.mlr.press/v80/pu18b.html}
\showURL{%
\tempurl}


\bibitem[\protect\citeauthoryear{Quirk, Mooney, and Galley}{Quirk
  et~al\mbox{.}}{2015}]%
        {quirk-etal-2015-language}
\bibfield{author}{\bibinfo{person}{Chris Quirk}, \bibinfo{person}{Raymond
  Mooney}, {and} \bibinfo{person}{Michel Galley}.}
  \bibinfo{year}{2015}\natexlab{}.
\newblock \showarticletitle{Language to Code: Learning Semantic Parsers for
  If-This-Then-That Recipes}. In \bibinfo{booktitle}{\emph{Proceedings of the
  53rd Annual Meeting of the Association for Computational Linguistics and the
  7th International Joint Conference on Natural Language Processing (Volume 1:
  Long Papers)}}. \bibinfo{publisher}{Association for Computational
  Linguistics}, \bibinfo{address}{Beijing, China}, \bibinfo{pages}{878--888}.
\newblock
\urldef\tempurl%
\url{https://doi.org/10.3115/v1/P15-1085}
\showDOI{\tempurl}


\bibitem[\protect\citeauthoryear{Rabinovich, Stern, and Klein}{Rabinovich
  et~al\mbox{.}}{2017}]%
        {rabinovich:2017:acllong}
\bibfield{author}{\bibinfo{person}{Maxim Rabinovich}, \bibinfo{person}{Mitchell
  Stern}, {and} \bibinfo{person}{Dan Klein}.} \bibinfo{year}{2017}\natexlab{}.
\newblock \showarticletitle{Abstract Syntax Networks for Code Generation and
  Semantic Parsing}. In \bibinfo{booktitle}{\emph{Proceedings of the 55th
  Annual Meeting of the Association for Computational Linguistics (Volume 1:
  Long Papers)}}. \bibinfo{publisher}{Association for Computational
  Linguistics}, \bibinfo{address}{Vancouver, Canada},
  \bibinfo{pages}{1139--1149}.
\newblock
\urldef\tempurl%
\url{https://doi.org/10.18653/v1/P17-1105}
\showDOI{\tempurl}


\bibitem[\protect\citeauthoryear{Ragan-Kelley, Barnes, Adams, Paris, Durand,
  and Amarasinghe}{Ragan-Kelley et~al\mbox{.}}{2013}]%
        {ragankelley:2013:acm}
\bibfield{author}{\bibinfo{person}{Jonathan Ragan-Kelley},
  \bibinfo{person}{Connelly Barnes}, \bibinfo{person}{Andrew Adams},
  \bibinfo{person}{Sylvain Paris}, \bibinfo{person}{Fr\'{e}do Durand}, {and}
  \bibinfo{person}{Saman Amarasinghe}.} \bibinfo{year}{2013}\natexlab{}.
\newblock \showarticletitle{Halide: A Language and Compiler for Optimizing
  Parallelism, Locality, and Recomputation in Image Processing Pipelines}.
\newblock \bibinfo{journal}{\emph{SIGPLAN Not.}} \bibinfo{volume}{48},
  \bibinfo{number}{6} (\bibinfo{date}{June} \bibinfo{year}{2013}),
  \bibinfo{pages}{519–530}.
\newblock
\showISSN{0362-1340}
\urldef\tempurl%
\url{https://doi.org/10.1145/2499370.2462176}
\showDOI{\tempurl}


\bibitem[\protect\citeauthoryear{Ratner, Bach, Ehrenberg, Fries, Wu, and
  R{\'e}}{Ratner et~al\mbox{.}}{2019}]%
        {ratner:2019:vldb}
\bibfield{author}{\bibinfo{person}{Alexander~J. Ratner},
  \bibinfo{person}{Stephen~H. Bach}, \bibinfo{person}{Henry~R. Ehrenberg},
  \bibinfo{person}{Jason~Alan Fries}, \bibinfo{person}{Sen Wu}, {and}
  \bibinfo{person}{C. R{\'e}}.} \bibinfo{year}{2019}\natexlab{}.
\newblock \showarticletitle{Snorkel: rapid training data creation with weak
  supervision}.
\newblock \bibinfo{journal}{\emph{The Vldb Journal}}  \bibinfo{volume}{29}
  (\bibinfo{year}{2019}), \bibinfo{pages}{709 -- 730}.
\newblock


\bibitem[\protect\citeauthoryear{Raychev, Bielik, and Vechev}{Raychev
  et~al\mbox{.}}{2016}]%
        {raychev:2016:oopsla}
\bibfield{author}{\bibinfo{person}{Veselin Raychev}, \bibinfo{person}{Pavol
  Bielik}, {and} \bibinfo{person}{Martin Vechev}.}
  \bibinfo{year}{2016}\natexlab{}.
\newblock \showarticletitle{Probabilistic Model for Code with Decision Trees}.
  In \bibinfo{booktitle}{\emph{Proceedings of the 2016 ACM SIGPLAN
  International Conference on Object-Oriented Programming, Systems, Languages,
  and Applications}} (Amsterdam, Netherlands) \emph{(\bibinfo{series}{OOPSLA
  2016})}. \bibinfo{publisher}{Association for Computing Machinery},
  \bibinfo{address}{New York, NY, USA}, \bibinfo{pages}{731–747}.
\newblock
\showISBNx{9781450344449}
\urldef\tempurl%
\url{https://doi.org/10.1145/2983990.2984041}
\showDOI{\tempurl}


\bibitem[\protect\citeauthoryear{Raychev, Vechev, and Krause}{Raychev
  et~al\mbox{.}}{2015}]%
        {raychev:2015:popl}
\bibfield{author}{\bibinfo{person}{Veselin Raychev}, \bibinfo{person}{Martin
  Vechev}, {and} \bibinfo{person}{Andreas Krause}.}
  \bibinfo{year}{2015}\natexlab{}.
\newblock \showarticletitle{Predicting Program Properties from "Big Code"}. In
  \bibinfo{booktitle}{\emph{Proceedings of the 42nd Annual ACM SIGPLAN-SIGACT
  Symposium on Principles of Programming Languages}} (Mumbai, India)
  \emph{(\bibinfo{series}{POPL '15})}. \bibinfo{publisher}{Association for
  Computing Machinery}, \bibinfo{address}{New York, NY, USA},
  \bibinfo{pages}{111–124}.
\newblock
\showISBNx{9781450333009}
\urldef\tempurl%
\url{https://doi.org/10.1145/2676726.2677009}
\showDOI{\tempurl}


\bibitem[\protect\citeauthoryear{Shin, Allamanis, Brockschmidt, and
  Polozov}{Shin et~al\mbox{.}}{2019}]%
        {shin2019program}
\bibfield{author}{\bibinfo{person}{Eui Chul~Richard Shin},
  \bibinfo{person}{Miltiadis Allamanis}, \bibinfo{person}{Marc Brockschmidt},
  {and} \bibinfo{person}{Alex Polozov}.} \bibinfo{year}{2019}\natexlab{}.
\newblock \showarticletitle{Program Synthesis and Semantic Parsing with Learned
  Code Idioms}. In \bibinfo{booktitle}{\emph{Advances in Neural Information
  Processing Systems 32: Annual Conference on Neural Information Processing
  Systems 2019, NeurIPS 2019, December 8-14, 2019, Vancouver, BC, Canada}},
  \bibfield{editor}{\bibinfo{person}{Hanna~M. Wallach}, \bibinfo{person}{Hugo
  Larochelle}, \bibinfo{person}{Alina Beygelzimer}, \bibinfo{person}{Florence
  d'Alch{\'{e}}{-}Buc}, \bibinfo{person}{Emily~B. Fox}, {and}
  \bibinfo{person}{Roman Garnett}} (Eds.). \bibinfo{pages}{10824--10834}.
\newblock
\urldef\tempurl%
\url{https://proceedings.neurips.cc/paper/2019/hash/cff34ad343b069ea6920464ad17d4bcf-Abstract.html}
\showURL{%
\tempurl}


\bibitem[\protect\citeauthoryear{Solar-Lezama}{Solar-Lezama}{2008}]%
        {10.5555/1714168}
\bibfield{author}{\bibinfo{person}{Armando Solar-Lezama}.}
  \bibinfo{year}{2008}\natexlab{}.
\newblock \emph{\bibinfo{title}{Program Synthesis by Sketching}}.
\newblock \bibinfo{thesistype}{Ph.D. Dissertation}. \bibinfo{address}{USA}.
\newblock Advisor(s) Bodik, Rastislav.
\newblock
\showISBNx{9781109097450}


\bibitem[\protect\citeauthoryear{Solar-Lezama, Tancau, Bodik, Seshia, and
  Saraswat}{Solar-Lezama et~al\mbox{.}}{2006}]%
        {10.1145/1168919.1168907}
\bibfield{author}{\bibinfo{person}{Armando Solar-Lezama},
  \bibinfo{person}{Liviu Tancau}, \bibinfo{person}{Rastislav Bodik},
  \bibinfo{person}{Sanjit Seshia}, {and} \bibinfo{person}{Vijay Saraswat}.}
  \bibinfo{year}{2006}\natexlab{}.
\newblock \showarticletitle{Combinatorial Sketching for Finite Programs}.
\newblock \bibinfo{journal}{\emph{SIGARCH Comput. Archit. News}}
  \bibinfo{volume}{34}, \bibinfo{number}{5} (\bibinfo{date}{Oct.}
  \bibinfo{year}{2006}), \bibinfo{pages}{404–415}.
\newblock
\showISSN{0163-5964}
\urldef\tempurl%
\url{https://doi.org/10.1145/1168919.1168907}
\showDOI{\tempurl}


\bibitem[\protect\citeauthoryear{Steedman}{Steedman}{1987}]%
        {Steedman1987CombinatoryGA}
\bibfield{author}{\bibinfo{person}{Mark Steedman}.}
  \bibinfo{year}{1987}\natexlab{}.
\newblock \showarticletitle{Combinatory grammars and parasitic gaps}.
\newblock \bibinfo{journal}{\emph{Natural Language \& Linguistic Theory}}
  \bibinfo{volume}{5} (\bibinfo{year}{1987}), \bibinfo{pages}{403--439}.
\newblock


\bibitem[\protect\citeauthoryear{Steedman and Baldridge}{Steedman and
  Baldridge}{2007}]%
        {Steedman2007COMBINATORYCG}
\bibfield{author}{\bibinfo{person}{Mark Steedman} {and} \bibinfo{person}{Jason
  Baldridge}.} \bibinfo{year}{2007}\natexlab{}.
\newblock \showarticletitle{Combinatory Categorical Grammar}.
\newblock


\bibitem[\protect\citeauthoryear{Subramanian, Bogin, Gupta, Wolfson, Singh,
  Berant, and Gardner}{Subramanian et~al\mbox{.}}{2020}]%
        {subramanian:2020:acl}
\bibfield{author}{\bibinfo{person}{Sanjay Subramanian}, \bibinfo{person}{Ben
  Bogin}, \bibinfo{person}{Nitish Gupta}, \bibinfo{person}{Tomer Wolfson},
  \bibinfo{person}{Sameer Singh}, \bibinfo{person}{Jonathan Berant}, {and}
  \bibinfo{person}{Matt Gardner}.} \bibinfo{year}{2020}\natexlab{}.
\newblock \showarticletitle{Obtaining Faithful Interpretations from
  Compositional Neural Networks}. In \bibinfo{booktitle}{\emph{ACL}}.
\newblock


\bibitem[\protect\citeauthoryear{Tran, Tran, Nguyen, Nguyen, and Nguyen}{Tran
  et~al\mbox{.}}{2019}]%
        {Tran_2019}
\bibfield{author}{\bibinfo{person}{Ngoc Tran}, \bibinfo{person}{Hieu Tran},
  \bibinfo{person}{Son Nguyen}, \bibinfo{person}{Hoan Nguyen}, {and}
  \bibinfo{person}{Tien Nguyen}.} \bibinfo{year}{2019}\natexlab{}.
\newblock \showarticletitle{Does BLEU Score Work for Code Migration?}
\newblock \bibinfo{journal}{\emph{2019 IEEE/ACM 27th International Conference
  on Program Comprehension (ICPC)}} (\bibinfo{date}{May} \bibinfo{year}{2019}).
\newblock
\showISBNx{9781728115191}
\urldef\tempurl%
\url{https://doi.org/10.1109/icpc.2019.00034}
\showDOI{\tempurl}


\bibitem[\protect\citeauthoryear{Waltz}{Waltz}{1978}]%
        {waltz:1978:acm}
\bibfield{author}{\bibinfo{person}{David~L. Waltz}.}
  \bibinfo{year}{1978}\natexlab{}.
\newblock \showarticletitle{An English Language Question Answering System for a
  Large Relational Database}.
\newblock \bibinfo{journal}{\emph{Commun. ACM}} \bibinfo{volume}{21},
  \bibinfo{number}{7} (\bibinfo{date}{July} \bibinfo{year}{1978}),
  \bibinfo{pages}{526–539}.
\newblock
\showISSN{0001-0782}
\urldef\tempurl%
\url{https://doi.org/10.1145/359545.359550}
\showDOI{\tempurl}


\bibitem[\protect\citeauthoryear{Wong and Mooney}{Wong and Mooney}{2006}]%
        {wong:2006:naacl}
\bibfield{author}{\bibinfo{person}{Yuk~Wah Wong} {and} \bibinfo{person}{Raymond
  Mooney}.} \bibinfo{year}{2006}\natexlab{}.
\newblock \showarticletitle{Learning for Semantic Parsing with Statistical
  Machine Translation}. In \bibinfo{booktitle}{\emph{Proceedings of the Human
  Language Technology Conference of the {NAACL}, Main Conference}}.
  \bibinfo{publisher}{Association for Computational Linguistics},
  \bibinfo{address}{New York City, USA}, \bibinfo{pages}{439--446}.
\newblock
\urldef\tempurl%
\url{https://www.aclweb.org/anthology/N06-1056}
\showURL{%
\tempurl}


\bibitem[\protect\citeauthoryear{Woods}{Woods}{1973}]%
        {woods:1973:ncce}
\bibfield{author}{\bibinfo{person}{W.~A. Woods}.}
  \bibinfo{year}{1973}\natexlab{}.
\newblock \showarticletitle{Progress in Natural Language Understanding: An
  Application to Lunar Geology}. In \bibinfo{booktitle}{\emph{Proceedings of
  the June 4-8, 1973, National Computer Conference and Exposition}} (New York,
  New York) \emph{(\bibinfo{series}{AFIPS '73})}.
  \bibinfo{publisher}{Association for Computing Machinery},
  \bibinfo{address}{New York, NY, USA}, \bibinfo{pages}{441–450}.
\newblock
\showISBNx{9781450379168}
\urldef\tempurl%
\url{https://doi.org/10.1145/1499586.1499695}
\showDOI{\tempurl}


\bibitem[\protect\citeauthoryear{Xiao, Dymetman, and Gardent}{Xiao
  et~al\mbox{.}}{2016}]%
        {xiao-etal-2016-sequence}
\bibfield{author}{\bibinfo{person}{Chunyang Xiao}, \bibinfo{person}{Marc
  Dymetman}, {and} \bibinfo{person}{Claire Gardent}.}
  \bibinfo{year}{2016}\natexlab{}.
\newblock \showarticletitle{Sequence-based Structured Prediction for Semantic
  Parsing}. In \bibinfo{booktitle}{\emph{Proceedings of the 54th Annual Meeting
  of the Association for Computational Linguistics (Volume 1: Long Papers)}}.
  \bibinfo{publisher}{Association for Computational Linguistics},
  \bibinfo{address}{Berlin, Germany}, \bibinfo{pages}{1341--1350}.
\newblock
\urldef\tempurl%
\url{https://doi.org/10.18653/v1/P16-1127}
\showDOI{\tempurl}


\bibitem[\protect\citeauthoryear{Xu, Vasilescu, and Neubig}{Xu
  et~al\mbox{.}}{2021}]%
        {xu2021inide}
\bibfield{author}{\bibinfo{person}{Frank~F. Xu}, \bibinfo{person}{Bogdan
  Vasilescu}, {and} \bibinfo{person}{Graham Neubig}.}
  \bibinfo{year}{2021}\natexlab{}.
\newblock \bibinfo{title}{In-IDE Code Generation from Natural Language: Promise
  and Challenges}.
\newblock
\newblock
\showeprint[arxiv]{2101.11149}~[cs.SE]


\bibitem[\protect\citeauthoryear{Ye, Zhou, Venkat, Marcus, Tatbul, Tithi,
  Hasabnis, Petersen, Mattson, Kraska, Dubey, Sarkar, and Gottschlich}{Ye
  et~al\mbox{.}}{2020}]%
        {ye2020misim}
\bibfield{author}{\bibinfo{person}{Fangke Ye}, \bibinfo{person}{Shengtian
  Zhou}, \bibinfo{person}{Anand Venkat}, \bibinfo{person}{Ryan Marcus},
  \bibinfo{person}{Nesime Tatbul}, \bibinfo{person}{Jesmin~Jahan Tithi},
  \bibinfo{person}{Niranjan Hasabnis}, \bibinfo{person}{Paul Petersen},
  \bibinfo{person}{Timothy Mattson}, \bibinfo{person}{Tim Kraska},
  \bibinfo{person}{Pradeep Dubey}, \bibinfo{person}{Vivek Sarkar}, {and}
  \bibinfo{person}{Justin Gottschlich}.} \bibinfo{year}{2020}\natexlab{}.
\newblock \bibinfo{title}{MISIM: A Novel Code Similarity System}.
\newblock
\newblock
\showeprint[arxiv]{2006.05265}~[cs.LG]


\bibitem[\protect\citeauthoryear{Yih, He, and Meek}{Yih et~al\mbox{.}}{2014}]%
        {yih:2014:acl}
\bibfield{author}{\bibinfo{person}{Wen-tau Yih}, \bibinfo{person}{Xiaodong He},
  {and} \bibinfo{person}{Christopher Meek}.} \bibinfo{year}{2014}\natexlab{}.
\newblock \showarticletitle{Semantic Parsing for Single-Relation Question
  Answering}. In \bibinfo{booktitle}{\emph{Proceedings of the 52nd Annual
  Meeting of the Association for Computational Linguistics (Volume 2: Short
  Papers)}}. \bibinfo{publisher}{Association for Computational Linguistics},
  \bibinfo{address}{Baltimore, Maryland}, \bibinfo{pages}{643--648}.
\newblock
\urldef\tempurl%
\url{https://doi.org/10.3115/v1/P14-2105}
\showDOI{\tempurl}


\bibitem[\protect\citeauthoryear{Yin and Neubig}{Yin and Neubig}{2017}]%
        {yin-neubig-2017-syntactic}
\bibfield{author}{\bibinfo{person}{Pengcheng Yin} {and} \bibinfo{person}{Graham
  Neubig}.} \bibinfo{year}{2017}\natexlab{}.
\newblock \showarticletitle{A Syntactic Neural Model for General-Purpose Code
  Generation}. In \bibinfo{booktitle}{\emph{Proceedings of the 55th Annual
  Meeting of the Association for Computational Linguistics (Volume 1: Long
  Papers)}}. \bibinfo{publisher}{Association for Computational Linguistics},
  \bibinfo{address}{Vancouver, Canada}, \bibinfo{pages}{440--450}.
\newblock
\urldef\tempurl%
\url{https://doi.org/10.18653/v1/P17-1041}
\showDOI{\tempurl}


\bibitem[\protect\citeauthoryear{Yin and Neubig}{Yin and Neubig}{2018}]%
        {yin:2018:emnlp}
\bibfield{author}{\bibinfo{person}{Pengcheng Yin} {and} \bibinfo{person}{Graham
  Neubig}.} \bibinfo{year}{2018}\natexlab{}.
\newblock \showarticletitle{TRANX: A Transition-based Neural Abstract Syntax
  Parser for Semantic Parsing and Code Generation}. In
  \bibinfo{booktitle}{\emph{EMNLP}}.
\newblock


\bibitem[\protect\citeauthoryear{Yu, Zhang, Yang, Yasunaga, Wang, Li, Ma, Li,
  Yao, Roman, Zhang, and Radev}{Yu et~al\mbox{.}}{2018}]%
        {yu:2018:emnlp}
\bibfield{author}{\bibinfo{person}{Tao Yu}, \bibinfo{person}{Rui Zhang},
  \bibinfo{person}{Kai Yang}, \bibinfo{person}{Michihiro Yasunaga},
  \bibinfo{person}{Dongxu Wang}, \bibinfo{person}{Zifan Li},
  \bibinfo{person}{James Ma}, \bibinfo{person}{Irene Li},
  \bibinfo{person}{Qingning Yao}, \bibinfo{person}{Shanelle Roman},
  \bibinfo{person}{Zilin Zhang}, {and} \bibinfo{person}{Dragomir Radev}.}
  \bibinfo{year}{2018}\natexlab{}.
\newblock \showarticletitle{{S}pider: A Large-Scale Human-Labeled Dataset for
  Complex and Cross-Domain Semantic Parsing and Text-to-{SQL} Task}. In
  \bibinfo{booktitle}{\emph{Proceedings of the 2018 Conference on Empirical
  Methods in Natural Language Processing}}. \bibinfo{publisher}{Association for
  Computational Linguistics}, \bibinfo{address}{Brussels, Belgium},
  \bibinfo{pages}{3911--3921}.
\newblock
\urldef\tempurl%
\url{https://doi.org/10.18653/v1/D18-1425}
\showDOI{\tempurl}


\bibitem[\protect\citeauthoryear{Zelle and Mooney}{Zelle and Mooney}{1996}]%
        {zelle:1996:aaai}
\bibfield{author}{\bibinfo{person}{John~M. Zelle} {and}
  \bibinfo{person}{Raymond~J. Mooney}.} \bibinfo{year}{1996}\natexlab{}.
\newblock \showarticletitle{Learning to Parse Database Queries Using Inductive
  Logic Programming}. In \bibinfo{booktitle}{\emph{Proceedings of the
  Thirteenth National Conference on Artificial Intelligence - Volume 2}}
  (Portland, Oregon) \emph{(\bibinfo{series}{AAAI'96})}.
  \bibinfo{publisher}{AAAI Press}, \bibinfo{pages}{1050–1055}.
\newblock
\showISBNx{026251091X}


\bibitem[\protect\citeauthoryear{Zettlemoyer and Collins}{Zettlemoyer and
  Collins}{2007}]%
        {zettlemoyer-collins-2007-online}
\bibfield{author}{\bibinfo{person}{Luke Zettlemoyer} {and}
  \bibinfo{person}{Michael Collins}.} \bibinfo{year}{2007}\natexlab{}.
\newblock \showarticletitle{Online Learning of Relaxed {CCG} Grammars for
  Parsing to Logical Form}. In \bibinfo{booktitle}{\emph{Proceedings of the
  2007 Joint Conference on Empirical Methods in Natural Language Processing and
  Computational Natural Language Learning ({EMNLP}-{C}o{NLL})}}.
  \bibinfo{publisher}{Association for Computational Linguistics},
  \bibinfo{address}{Prague, Czech Republic}, \bibinfo{pages}{678--687}.
\newblock
\urldef\tempurl%
\url{https://www.aclweb.org/anthology/D07-1071}
\showURL{%
\tempurl}


\bibitem[\protect\citeauthoryear{Zettlemoyer and Collins}{Zettlemoyer and
  Collins}{2005}]%
        {zettlemoyer:2005:auai}
\bibfield{author}{\bibinfo{person}{Luke~S. Zettlemoyer} {and}
  \bibinfo{person}{Michael Collins}.} \bibinfo{year}{2005}\natexlab{}.
\newblock \showarticletitle{Learning to Map Sentences to Logical Form:
  Structured Classification with Probabilistic Categorial Grammars}. In
  \bibinfo{booktitle}{\emph{Proceedings of the Twenty-First Conference on
  Uncertainty in Artificial Intelligence}} (Edinburgh, Scotland)
  \emph{(\bibinfo{series}{UAI'05})}. \bibinfo{publisher}{AUAI Press},
  \bibinfo{address}{Arlington, Virginia, USA}, \bibinfo{pages}{658–666}.
\newblock
\showISBNx{0974903914}


\bibitem[\protect\citeauthoryear{Zhang, Yang, Baghdadi, Kamil, Shun, and
  Amarasinghe}{Zhang et~al\mbox{.}}{2018}]%
        {zhang:2018:acm}
\bibfield{author}{\bibinfo{person}{Yunming Zhang}, \bibinfo{person}{Mengjiao
  Yang}, \bibinfo{person}{Riyadh Baghdadi}, \bibinfo{person}{Shoaib Kamil},
  \bibinfo{person}{Julian Shun}, {and} \bibinfo{person}{Saman Amarasinghe}.}
  \bibinfo{year}{2018}\natexlab{}.
\newblock \showarticletitle{GraphIt: A High-Performance Graph DSL}.
\newblock \bibinfo{journal}{\emph{Proc. ACM Program. Lang.}}
  \bibinfo{volume}{2}, \bibinfo{number}{OOPSLA}, Article
  \bibinfo{articleno}{121} (\bibinfo{date}{Oct.} \bibinfo{year}{2018}),
  \bibinfo{numpages}{30}~pages.
\newblock
\urldef\tempurl%
\url{https://doi.org/10.1145/3276491}
\showDOI{\tempurl}


\bibitem[\protect\citeauthoryear{Zhong, Xiong, and Socher}{Zhong
  et~al\mbox{.}}{2017}]%
        {zhong:2018:arxiv}
\bibfield{author}{\bibinfo{person}{Victor Zhong}, \bibinfo{person}{Caiming
  Xiong}, {and} \bibinfo{person}{Richard Socher}.}
  \bibinfo{year}{2017}\natexlab{}.
\newblock \bibinfo{title}{Seq2{SQL}: Generating Structured Queries From Natural
  Language Using Reinforcement Learning}.
\newblock
\newblock
\showeprint[arxiv]{1709.00103}~[cs.CL]


\end{thebibliography}

\appendix

\end{document}